\documentclass[aps,pra,twocolumn,groupedaddress]{revtex4-1}

\pdfoutput=1

\usepackage{eurosym}
\usepackage{amsfonts}
\usepackage{amsmath}
\usepackage{amssymb,epsf}
\usepackage{color}

\usepackage{epstopdf}
\usepackage{graphicx}
\usepackage{float}
\usepackage{subfig}
\usepackage{hyperref}
\usepackage{mathtools}

\definecolor{aogreen}{rgb}{0.0, 0.5, 0.0}

\def\ketm#1{  \left\vert  #1   \right\rangle   }

\begin{document}

\title{Fidelity susceptibility near topological phase transitions in quantum walks}

\author{S. Panahiyan $^{1,2,3}$}
\email{email address: shahram.panahiyan@uni-jena.de}
\author{W. Chen $^{4}$}
\email{email address: wchen@puc-rio.br}
\author{S. Fritzsche $^{1,2,3}$ }
\email{email address: s.fritzsche@gsi.de}
\affiliation{$^1$Helmholtz-Institut Jena, Fr\"{o}belstieg 3, D-07743 Jena, Germany  \\
	$^2$GSI Helmholtzzentrum f\"{u}r Schwerionenforschung, D-64291 Darmstadt, Germany \\
	$^3$Theoretisch-Physikalisches Institut, Friedrich-Schiller-University Jena, D-07743 Jena, Germany\\
	Department of Physics, PUC-Rio, 22451-900 Rio de Janeiro, Brazil}

\date{\today}

\begin{abstract}

The notion of fidelity susceptibility, introduced within the context of quantum metric tensor, has been an important quantity to characterize the criticality near quantum phase transitions. We demonstrate that for topological phase transitions in Dirac models, provided the momentum space is treated as the manifold of the quantum metric, the fidelity susceptibility coincides with the curvature function whose integration gives the topological invariant. Thus the quantum criticality of the curvature function near a topological phase transition also describes the criticality of the fidelity susceptibility, and the correlation length extracted from the curvature function also gives a momentum scale over which the fidelity susceptibility decays. To map out the profile and criticality of the fidelity susceptibility, we turn to quantum walks that simulate one-dimensional class BDI and two-dimensional class D Dirac models, and demonstrate their accuracy in capturing the critical exponents and scaling laws near topological phase transitions.
\end{abstract}

\maketitle

\section{Introduction}

In topologically ordered materials, the change of topological invariant caused by tuning a certain system parameter signifies a topological phase transition. In an attempt to draw analogy with the usual second-order quantum phase transition, a recently emerged scenario is to characterize the quantum criticality near the topological phase transition through investigating the curvature function, defined as the function whose momentum space integration gives the topological invariant\cite{Chen2017,Chen19_book_chapter,Chen20191,Molignini19}. From the generic critical behavior of the curvature function, the critical exponents and scaling laws can be extracted, and the Fourier transform of the curvature function is proposed to be the characteristic correlation function for topological materials. Because the curvature function is a purely geometric object that is not limited to a specific dimension or symmetry class\cite{Schnyder08,Ryu10,Chiu16}, this scenario has successfully described the criticality in a wide range of topological materials including almost all prototype noninteracting models\cite{Chen20191}, weakly\cite{Chen18_weakly_interacting} and strongly interacting models\cite{Kourtis17}, as well as periodically driven\cite{Molignini18,Molignini20} and multicritical systems\cite{Rufo19,Abdulla20}. In addition, a curvature renormalization group (CRG) approach has been proposed based on the divergence of the curvature function\cite{Chen1,Chen2}, which is shown to be particularly powerful in solving otherwise tedious interacting or mult-parameter systems\cite{Chen18_weakly_interacting,Kourtis17,Niewenburg,Molignini19}. 

On the other hand, another important quantity that has been proposed to characterize the criticality near quantum phase transitions in general is the fidelity susceptibility\cite{You07,Zanardi07}. Formulated within the framework of quantum metric tensor, the fidelity susceptibility measures how a tuning parameter makes a quantum state deviate from itself, and usually diverges as the system approaches the critical point. In addition, the scaling behavior of the fidelity susceptibility near the critical point has been investigated in a variety of correlated models\cite{Gu08,Yang08,Albuquerque10,Gu10}. Given that this aspect of fidelity susceptibility can be broadly applied to any type of quantum phase transitions, it is intriguing to ask how the fidelity susceptibility manifests in topological phase transitions, whether it displays a particular scaling behavior, and how it is related to the aforementioned scenario based on the curvature function. 

In this paper, we formulate the fidelity susceptibility for topological phase transitions within the framework of Dirac models, which are low energy effective theories for a wide range of topological materials. We observe that the fidelity ssuceptibility has a meaningful intepretation if one treats the momentum space as a manifold to construct the quantum metric. For systems where the curvature function is the Berry connection or Berry curvature, the determinant of the quantum metric tensor coincides with the square of the curvature function. As a result, the fidelity susceptibility shares the same critical behavior as the curvature function, and so inherits its critical exponent and scaling laws, and moreover decays with a characteristic momentum scale resulted from the correlation length.

We further address the issue of pratically simulating the criticality of curvature function and fidelity susceptibility. For this purpose, we turn to the quantum walks, which are proposed to be universal primitives \cite{Lovett} that can simulate a variety of quantum systems and phenomena \cite{Mohseni,Vakulchyk}, including topological materials \cite{Kitagawa}. The flexibility and controllability of the quantum walks help to simulate these topological phases\cite{Panahiyan2019}, which include all symmetry classes and edge states in one- (1D) and two-dimensional (2D) systems \cite{Kitagawa,Panahiyan2019,Asboth,Obuse,Chen,Panahiyan2020-1}, some others in three-dimension (3D) \cite{Panahiyan2020-2}, as well as to directly probe the topological invariants \cite{Ramasesh}. The topological phase transitions \cite{Rakovszky} and the possibility to invoke bulk-boundary correspondence have also been addressed for quantum walks. From experimental point of view, the existence of a robust edge state for the simulated topological phases was reported \cite{KitagawaExp} and it was shown that experimental realizations of the quantum walk can be employed to investigate topological phenomena in both 1D and 2D \cite{Cardano,Cardano2017,Barkhofen,Flurin,Zhan,Xiao,Wang2018,Wang,Nitsche,Errico,Xu}. 
This list of encouraging results is enriched by our analysis that, by drawing analogy with perdiodically driven systems\cite{Molignini18,Molignini20}, clarifies the stroboscopic curvature functions for two specific cases of quantum walks in 1D and 2D. The extracted critical exponents indicate that these quantum walks faithfully reproduce the desired critical behavior, and as simulators belong to the same universality classes as their topological insulator counterparts. A correlation function that measures the overlap of stroboscopic Wannier states is proposed, and the convenience of CRG in solving multi-parameter quantum walks is elaborated.

The structure of the paper is organized in the following manner. In Sec.~\ref{sec:quantum_criticality_TPT}, we give an overview of the curvature function scenario, including the generic critical behavior, critical exponents, and the CRG approach. We then calculate the fidelity susceptibility in 1D and 2D Dirac models, and show explicitly its coincidence with the Berry connection and Berry curvature. In Sec.~\ref{sec:quantum_walks}, we turn to quantum walks in two specific cases that simulate the Berry curvature and Berry connection, and extract the critical exponents and scaling laws that are consistent with the Dirac models. Section \ref{sec:conclusion} gives a summary and outlook of these results. 

\section{Quantum criticality near topological phase transitions \label{sec:quantum_criticality_TPT}}

\subsection{Generic critical behavior \label{sec:generic_critical_behavior}}

The eigenstates of a $D$ dimensional noninteracting topological material can in general be parameterized by two distinctive set of parameters; $\boldsymbol k$ and $\boldsymbol M$ in which $\boldsymbol k$ are momenta in $D$ dimension and $\boldsymbol M$ are tunable parameters in the Hamiltonian. Different topological phases are characterized by quantized integers known as topological invariants, which are generally obtained through integration of a curvature function over the first Brillouin zone (BZ)\cite{Chen2017,Chen20191,Molignini19}
\begin{eqnarray}
C(M) &=&  \int_{BZ} F(\boldsymbol k,\boldsymbol M)  \frac{d^d  \boldsymbol k}{(2 \pi)^d}, \label{TPI}
\end{eqnarray} 
in which $F(\boldsymbol k,\boldsymbol M)$ is referred to as the curvature function. Different topological phases are separated by boundaries that define topological phase transitions. As $\boldsymbol M$ crosses the critical point $\boldsymbol M_{c}$, the topological invariant jumps from one integer to another, accompanied by a gap-closing in the energy spectrum.  

The precise definition of the curvature function depends on the dimensionality and symmetries of the system under consideration\cite{Chen19_book_chapter,Chen20191}. It is generally an even function $F(\boldsymbol k_{c}+\delta \boldsymbol k)=F(\boldsymbol k_{c}-\delta \boldsymbol k)$ around a certain momentum $\boldsymbol k_{c}$ in the BZ, and hence well described by an Ornstein-Zernike form
\begin{align}
F(k_{c}+\delta k,\boldsymbol M) &=  \frac{F(k_{c},\boldsymbol M)}{ 1 \pm \xi^2 \delta k^2}, \label{curv-11D}
\\
F(\boldsymbol k_{c}+\delta \boldsymbol k,\boldsymbol M) &= \frac{F(\boldsymbol k_{c},\boldsymbol M)}{ (1 \pm \xi^2_{x} \delta k^2_{x}) (1 \pm \xi^2_{y} \delta k^2_{y})}, \label{curv-22D}
\end{align}
in 1D and 2D, respectively, in which $\xi$, $\xi_{x}$ and $\xi_{y}$ are width of the peak and they are characteristic length scales. The key ingredient of the curvature function as a tool to investigate topological phase transition is its varying nature as $\boldsymbol M$ changes. In other words, the topological invariant remains fixed for specific region of $\boldsymbol M$ whereas the profile of the curvature function varies. This variation enables us to characterize the critical behavior of the system with curvature function and extract correlation function, critical exponents, length scale, and validate a scaling law\cite{Chen2017,Chen19_book_chapter,Chen20191}. 

The critical behavior of the curvature function in the majority of topological materials is described by the narrowing and flipping of the Lorentian peak of the curvature function as the system approaches the two sides of the critical point $\boldsymbol M^{+}_{c}$ and $\boldsymbol M^{-}_{c}$
\begin{eqnarray}
&&\lim\limits_{\boldsymbol M \rightarrow \boldsymbol M^{+}_{c}} F(\bf k_{c},\boldsymbol M) = - \lim\limits_{\alpha \rightarrow \boldsymbol M^{-}_{c}} F(\bf k_{c},\boldsymbol M) = \pm \infty,
\\
&&\lim\limits_{\boldsymbol M \rightarrow \boldsymbol M_{c}} \xi = \infty,
\label{FkM_xi_critical_behavior}
\end{eqnarray}
which are also found to be true for quantum walks, as we demonstrate in later sections. These divergencies suggest that the critical behavior of $F(\bf k_{c},\bf M)$ and $\xi$ can be described by
\begin{eqnarray}
F(\bf k_{c},\boldsymbol M)\propto |\boldsymbol M-\boldsymbol M_{c}|^{-\gamma} \text{,                } \xi \propto |\boldsymbol M-\boldsymbol M_{c}|^{-\nu}
\end{eqnarray}
in which $\gamma$ and $\nu$ are critical exponents that satisfy a scaling law $\gamma= D\nu$ originated from the conservation of the topological invariant \cite{Chen2017,Chen20191,Chen19_book_chapter}. The physical meaning of these exponents become transparent through the notion of correlation functions, which is introduced by considering the Wannier states constructed from the Bloch state of the Hamiltonian
\begin{eqnarray}
|{\boldsymbol R}\rangle=\frac{1}{N}\sum_{\boldsymbol k}e^{i{\boldsymbol k({\hat r}-R)}}|\psi_{\boldsymbol k-}\rangle.
\end{eqnarray}
in which $|\psi_{\boldsymbol k-}\rangle$ is the lower eigenstate of the Hamiltonian. The correlation function is proposed to be the Fourier transform of the curvature function, which generally measures the overlap of Wannier states at the origin $|\boldsymbol 0\rangle$ and at $|\boldsymbol R\rangle$ sandwiched by a certain position operator. In 1D case where the curvature function is the Berry connection, the Wannier state correlation function reads
\begin{eqnarray}
\tilde{F}_{1D}(R)&=&\int_{0}^{2\pi}\frac{dk_x}{2\pi}F(k_x,\boldsymbol M) e^{ik_xR}   \notag
\\
&=&\int_{0}^{2\pi}\frac{dk}{2\pi}\langle\psi_{-}|i\partial_{k_x}|\psi_{-}\rangle e^{ik_xR}
  \notag
\\
&=&\langle 0|{\hat r}|R\rangle=\int dr\,r\,W^{\ast}(r)W(r-R),
\label{1D_Wannier_correlation}
\end{eqnarray}
where $\langle r|R\rangle=W(r-R)$ is the Wannier function centering at the home cell $R$, and ${\hat r}$ is the position operator. By replacing the curvature function with $\eqref{curv-11D}$, one can show that the Wannier state correlation function decays with the length scale $\xi$, and hence $\xi$ can be interpreted as the correlation length in this problem, assigned with the critical exponent $\nu$ as in the convention of statistical mechanics. The same can be done for 2D case where the curvature function is the Berry curvature
\begin{eqnarray*}
&&\tilde{F}_{2D}({\boldsymbol R})=\int\frac{d^{2}{\bf k}}{(2\pi)^{2}}F({\boldsymbol k},\boldsymbol M)=
\nonumber \\
&&\int\frac{d^{2}{\boldsymbol k}}{(2\pi)^{2}}\left\{\partial_{k_{x}}\langle\psi_{\boldsymbol k-}|i\partial_{k_{y}}|\psi_{\boldsymbol k-}\rangle-
\partial_{k_{y}}\langle\psi_{\boldsymbol k-}|i\partial_{k_{x}}|\psi_{\boldsymbol k-}\rangle\right\}e^{i{\boldsymbol k\cdot R}}
\nonumber \\
&&=-i\langle{\boldsymbol R}|(R^{x}{\hat y}-R^{y}{\hat x})|{\bf 0}\rangle
\nonumber \\
&&=-i\int d^{2}{\boldsymbol r}(R^{x}y-R^{y}x)W^{\ast}({\boldsymbol r}-{\boldsymbol R})W({\boldsymbol r}),
\label{Wannier_correlation_2D}
\end{eqnarray*} 
in which $\langle{\boldsymbol r}|{\boldsymbol R}\rangle=W({\boldsymbol r}-{\boldsymbol R})$ is the Wannier function. In the following sections, we will elaborate explicitly that topological quantum walks also fit into this scheme of curvature-based quantum criticality.

\subsection{Curvature renormalization group approach}

In systems whose topology is controlled by multiple tuning parameters ${\boldsymbol M}=(M_{1},M_{2}...)$, a curvature renormalization group (CRG) has been proposed to efficiently capture the topological phase transitions in the multi-dimensional parameter space\cite{Chen1,Chen2}. The approach is based on the iterative mapping ${\boldsymbol M}\rightarrow{\boldsymbol M}'$ that satisfies
\begin{equation}
F({\boldsymbol k}_{0}+\delta {\boldsymbol k},{\boldsymbol M})=F({\boldsymbol k}_{0},{\boldsymbol M}'),
\end{equation}
where $\delta {\boldsymbol k}=\delta k{\hat{\boldsymbol k}}_{s}$ is a small deviation away from the high symmetry point (HSP) ${\boldsymbol k}_{0}$ along the scaling direction ${\hat{\boldsymbol k}}_{s}$. Expanding the scaling equation up to leading order yields the generic RG equation
\begin{equation}
\frac{dM_{i}}{d\ell}=\frac{M_{i}^{\prime}-M_{i}}{\delta k^{2}}=\frac{1}{2}\frac{({\boldsymbol\nabla}\cdot{\boldsymbol {\hat k}}_{s})^{2}F({\boldsymbol k},{\boldsymbol M})|_{\boldsymbol k=k_{0}}}{\partial_{M_{i}}F({\boldsymbol k}_{0},{\boldsymbol M})}
\label{RG_eq_derivative}
\end{equation}
Numerically, the right-hand side of the above equation can be evaluated conveniently by
\begin{equation}
\frac{dM_{i}}{d\ell}=\frac{F({\boldsymbol k}_{0}+\Delta k{\boldsymbol{\hat k}}_{s},{\boldsymbol M})-F({\boldsymbol k}_{0},{\boldsymbol M})}{F({\boldsymbol k}_{0},{\boldsymbol M}+\Delta M_{i}{\hat{\boldsymbol M}}_{i})-F({\boldsymbol k}_{0},{\boldsymbol M})},
\label{RG_eq_numerical}
\end{equation}
where $\Delta k$ is a small deviation away from the HSP in momentum space, and $\Delta M_{i}$ is a small interval in the parameter space along the ${\hat{\boldsymbol M}}_{i}$ direction. This numerical interpretation serves as a great advantage over the integration of topological invariant in Eq.~(\ref{TPI}), since for a given $\boldsymbol M$, one only needs to calculate the curvature function at three points $F({\boldsymbol k}_{0}+\Delta k{\boldsymbol{\hat k}}_{s},{\boldsymbol M})$, $F({\boldsymbol k}_{0},{\bf M})$ and $F({\boldsymbol k}_{0},{\boldsymbol M}+\Delta M_{i}{\hat{\boldsymbol M}}_{i})$ to obtain the RG flow along the ${\hat{\boldsymbol M}}_{i}$ direction, and hence a powerful tool to capture the topological phase transitions in the vast ${\boldsymbol M}$ parameter space. The efficiency of this method has been demonstrated in a great variety of systems, and in the present work we aim to demonstrate its feasibility for quantum walks.

\subsection{Fidelity susceptibility near topological phase transitions}

In this section, we elaborate that within the context of Dirac models, the fidelity susceptibility near a topological phase transition has the same critical behavior as the curvature function. For completeness, we first give an overview of the fidelity susceptibility formulated under the notion of quantum geometric tensor\cite{Zanardi07,You07,Gu10}. Our aim is to calculate the fidelity of the eigenstates of a given Hamiltonian under one or multiple tuning parameters
 \begin{eqnarray}
&&H(\mu)=H_{0}+\mu\,H_{I},\;\;\;H(\mu)|\psi_{n}(\mu)\rangle=E_{n}|\psi_{n}(\mu)\rangle,
\nonumber \\
&&\sum_{n}|\psi_{n}(\mu)\rangle\langle \psi_{n}(\mu)|=I,
\end{eqnarray}
where $\mu=\left\{\mu_{a}\right\}$ with $a=1,2...\Lambda$ is a set of tuning parameters that form a $\Lambda$-dimensional manifold. For two eigenstates that are very close in the parameter space, the fidelity is the module of the product of the two eigenstates
\begin{equation}
|\langle\psi(\mu)|\psi(\mu+\delta\mu)\rangle|= 1-\sum_{ab}\frac{1}{2}g_{ab}\delta\mu_{a}\delta\mu_{b},
\end{equation}
which defines the the quantum metric tensor
\begin{eqnarray}
g_{ab}=\frac{1}{2}\langle\partial_{a}\psi|\partial_{b}\psi\rangle
+\frac{1}{2}\langle\partial_{b}\psi|\partial_{a}\psi\rangle
-\langle\partial_{a}\psi|\psi\rangle\langle\psi|\partial_{b}\psi\rangle,
\nonumber \\
\label{gab_multiple_lambda}
\end{eqnarray} 
as a measure of the distance between the quantum states in the $\left\{\mu_{a}\right\}$ manifold. The quantum metric tensor is the real part $g_{ab}={\rm Re}T_{ab}$ of the more general quantum geometric tensor 
\begin{eqnarray}
T_{ab}=\langle\partial_{a}\psi|\partial_{b}\psi\rangle-\langle\partial_{a}\psi|\psi\rangle\langle\psi|\partial_{b}\psi\rangle.
\label{Tab_definition}
\end{eqnarray}
whose imaginary part is the Berry curvature times $-1/2$
\begin{eqnarray}
{\rm Im}T_{ab}=-\frac{1}{2}\left[\partial_{a}\langle\psi_{-}|i\partial_{b}|\psi_{-}\rangle
-\partial_{b}\langle\psi_{-}|i\partial_{a}|\psi_{-}\rangle\right].
\end{eqnarray}
It can be easily shown that the quantum geometric tensor is invariant under local gauge transformation $|\psi_{-}\rangle\rightarrow e^{i\varphi}|\psi_{-}\rangle$, and so are $g_{ab}$ and the Berry curvature. Hence these quantities are measurables, as have been demonstrated in various systems\cite{Abanin13,Jotzu14,Duca15,Tan19}.

\subsubsection{1D topological insulators \label{sec:1D_Dirac_model}}

We now consider the quantum metric tensor of the eigenstates of a $2\times 2$ Dirac Hamiltonian that has only two components
\begin{eqnarray}
H=d_{1}\sigma_{1}+d_{2}\sigma_{2},
\label{1D_Dirac_d1d2}
\end{eqnarray}
which are relevent to several classes of 1D topological insulators\cite{Schnyder08,Ryu10,Chiu16,Chen20191}. The eigenstates and eigenenergies are, in a specific gauge choice,
\begin{eqnarray}
|\psi_{\pm}\rangle=\frac{1}{\sqrt{2}d}\left(\begin{array}{c}
\pm d \\
d_{1}+id_{2}
\end{array}
\right),\;\;\;E_{\pm}=\pm d,
\label{1D_eigenstates}
\end{eqnarray}
where $d=\sqrt{d_{1}^{2}+d_{2}^{2}}$. Suppose each component of the ${\bf d}$-vector is a function of a certain tuning parameter $k$ (what precisely is $k$ is unimportant at this stage), then the quantum metric tensor in Eq.~(\ref{gab_multiple_lambda}) reads
\begin{eqnarray}
g_{kk}&=&\langle\partial_{k}\psi_{-}|\partial_{k}\psi_{-}\rangle-\langle\partial_{k}\psi_{-}|\psi_{-}\rangle\langle\psi_{-}|\partial_{k}\psi_{-}\rangle
\nonumber \\
&=&\left[\frac{d_{1}\partial_{k}d_{2}-d_{2}\partial_{k}d_{1}}{2d^{2}}\right]^{2}
=\left[\langle\psi_{-}|i\partial_{k}|\psi_{-}\rangle\right]^{2}
\nonumber \\
&=&\frac{1}{4}({\hat{\bf d}}\times\partial_{k}{\hat{\bf d}})_{z}^{2}=\frac{1}{4}\partial_{k}{\hat{\bf d}}\cdot\partial_{k}{\hat{\bf d}},
\label{1D_gkk_Berry2}
\end{eqnarray}
which is equal to the square of the Berry connection $\langle\psi_{-}|i\partial_{k}|\psi_{-}\rangle$ in this gauge (it should be remined that $g_{kk}$ is gauge invariant but the Berry connection is not), and ${\hat{\bf e}}_{k}=\partial_{k}{\hat{\bf d}}/2$ plays the role of vielbein. Moreover, in this case that there is only one tuning parameter $k$, the quantum metric tensor is also equal to the fidelity susceptibility defined from\cite{You07,Zanardi07} 
\begin{eqnarray}
\langle\psi_{-}(k)|\psi_{-}(k+\delta k)\rangle=1-\frac{\delta k^{2}}{2}\chi_{F}=1-\frac{\delta k^{2}}{2}g_{kk}.
\label{1D_fidelity_sus_definition}
\end{eqnarray}

We proceed to consider the physically meaningful Dirac model relevant to the low energy theory near the HSP $k=0$ of topological insulators
\begin{eqnarray}
d_{1}=M,\;\;\;d_{2}=k,
\end{eqnarray}
where $M$ is the mass and $k$ is the momentum. Our observation is that the quantum metric has a meaningful interpretation if we treat the momentum space as a manifold and construct the metric between $|\psi_{-}(k)\rangle$ and $|\psi_{-}(k+\delta k)\rangle$. Using Eqs.~(\ref{1D_gkk_Berry2}) and (\ref{1D_fidelity_sus_definition}), this gives
\begin{eqnarray}
&&\langle\psi_{-}|i\partial_{k}|\psi_{-}\rangle=-\frac{M}{2(M^{2}+k^{2})},
\nonumber \\
&&\chi_{F}=\frac{M^{2}}{4(M^{2}+k^{2})^{2}}.
\end{eqnarray}
At the HSP $k=0$, these quantities diverge with the mass term $M$ like
\begin{eqnarray}
&&\langle\psi_{-}|i\partial_{k}|\psi_{-}\rangle|_{k=0}\propto |M|^{-1}=|M|^{-\gamma},
\nonumber \\
&&\chi_{F}|_{k=0}\propto |M|^{-2}=|M|^{-2\gamma}.\;\;\;
\label{chiF_div_1D}
\end{eqnarray}
Thus the divergence of the Berry connection and that of the fidelity susceptibility near the topological phase transition $M\rightarrow 0$ are basically described by the same critical exponent $\gamma$. This justifies the usage of exponent $\gamma$ for the Berry connection at $k=0$ which is conventionally assigned to the susceptibility. Notice that in topological phase transitions, we treat $M$ as the tuning parameter, but to extract the quantum metric we treat $k$ as the tuning parameter.

Equation (\ref{1D_gkk_Berry2}) has another significant implication on the differential geometry of the manifold. Because the determinant of the quantum metric is the quantum metric itself $\det g_{kk}=g_{kk}\equiv g$, it implies that the integration $I=\int\phi(k)\sqrt{g}dk$ of any function $\phi(k)$ over the manifold is associated with the line element $\sqrt{g}dk=|\langle\psi_{-}|i\partial_{k}|\psi_{-}\rangle|dk$ given by the absolute value of the Berry connection. Therefore the total length of the 1D manifold is
\begin{eqnarray}
L=\int\sqrt{g}dk=\int|\langle\psi_{-}|i\partial_{k}|\psi_{-}\rangle|dk.
	\label{length_1D_manifold}
\end{eqnarray}
In the topologically nontrivial phase of a varieties of 1D topological insulators, such as the Su-Schrieffer-Heeger model\cite{Chen2017,Chen19_book_chapter}, the Berry connection is often positive everywhere on the manifold $\langle\psi_{-}|i\partial_{k}|\psi_{-}\rangle=|\langle\psi_{-}|i\partial_{k}|\psi_{-}\rangle|$ (this also occurs in our 1D quantum walk at some parameters, as discussed in Sec.~\ref{sec:quantum_walks}). In this case, the total length of the manifold is equal to the topological invariant $L={\cal C}$, and hence remains a integer. In contrast, for the topologically trivial phase where the Berry connection is positive in some regions and negative in some other, this equivalence is not guaranteed and the length $L$ should be calculated by Eq.~(\ref{length_1D_manifold}).

\subsubsection{2D time-reversal breaking topological insulators}

We now turn to the 2D Dirac Hamiltonian that has all three components 
\begin{eqnarray}
H=d_{1}\sigma_{1}+d_{2}\sigma_{2}+d_{3}\sigma_{3},
\end{eqnarray}
which has eigenstates and eigenenergies
\begin{eqnarray}
|\psi_{\pm}\rangle=\frac{1}{\sqrt{2d(d\pm d_{3})}}\left(\begin{array}{c}
d_{3}\pm d \\
d_{1}+id_{2}
\end{array}\right),\;\;\;E_{\pm}=\pm d,
\end{eqnarray}
where $d=\sqrt{d_{1}^{2}+d_{2}^{2}+d_{3}^{2}}$. Since the two eigenstates form a complete set $I=|\psi_{+}\rangle\langle\psi_{+}|+|\psi_{-}\rangle\langle\psi_{-}|$, we can compute the quantum geometric tensor $T_{ab}$ in Eq.~(\ref{Tab_definition}) by assuming each component of the ${\bf d}$-vector is a function of $\left\{\mu_{a}\right\}$, 
\begin{eqnarray}
&&T_{ab}=\langle\partial_{a}\psi_{-}|\psi_{+}\rangle\langle\psi_{+}|\partial_{b}\psi_{-}\rangle
\nonumber \\
&&=\frac{1}{4d^{2}(d_{1}^{2}+d_{2}^{2})}\left(-d_{3}\partial_{a}d+d\partial_{a}d_{3}-id_{1}\partial_{a}d_{2}+id_{2}\partial_{a}d_{1}\right)
\nonumber \\
&&\times\left(-d_{3}\partial_{b}d+d\partial_{b}d_{3}+id_{1}\partial_{b}d_{2}-id_{2}\partial_{b}d_{1}\right),
\end{eqnarray}
whose real and imaginary parts are
\begin{eqnarray}
&&{\rm Re}T_{ab}=g_{ab}=\frac{1}{4}\partial_{a}{\hat{\bf d}}\cdot\partial_{b}{\hat{\bf d}}
\nonumber \\
&&=\frac{1}{4d^{2}}\left\{\partial_{a}d_{1}\partial_{b}d_{1}+\partial_{a}d_{2}\partial_{b}d_{2}
+\partial_{a}d_{3}\partial_{b}d_{3}-\partial_{a}d\partial_{b}d\right\},
\nonumber \\
&&{\rm Im}T_{ab}=-\frac{1}{2}\Omega_{ab}=-\frac{1}{4}{\hat{\bf d}}\cdot\left(\partial_{a}{\hat{\bf d}}\times\partial_{b}{\hat{\bf d}}\right)
\nonumber \\
&&=-\frac{1}{4d^{3}}\epsilon^{ijk}d_{i}\partial_{a}d_{j}\partial_{b}d_{k}.
\label{2D_Regab_Imgab}
\end{eqnarray}
One sees that ${\hat{\bf e}}_{a}=\partial_{a}{\hat{\bf d}}/2$ plays the role of vielbein according the the definition $g_{ab}={\hat{\bf e}}_{a}\cdot{\hat{\bf e}}_{b}$, and $\Omega_{ab}$ is the Berry curvature whose integration over the 2D manifold gives the skyrmion number of the ${\hat{\bf d}}$ vector. Moreover, the determinant of the quantum metric tensor coincides with the square of the Berry curvature
\begin{eqnarray}
g=\det g_{ab}=\frac{1}{4}\Omega_{xy}^{2}\equiv \chi_{F},
\label{chiF_div_2D}
\end{eqnarray}
a result very similar to Eq.~(\ref{1D_gkk_Berry2}), which suggests the determinant $g\equiv \chi_{F}$ as the representative fidelity susceptibility in this 2D problem.
	
Similar to that discussed before and after Eq.~(\ref{length_1D_manifold}), the determinant $g$ gives the area element $\sqrt{g}d^{2}{\bf k}$ of the integration of any function over the 2D manifold. Thus the total area of the manifold reads
\begin{eqnarray}
A=\int\sqrt{g}\,d^{2}{\bf k}=\frac{1}{2}\int|\Omega_{xy}|d^{2}{\bf k}.
\end{eqnarray}
Thus for the cases that the Berry curvature is positive everywhere on the manifold $\Omega_{xy}=|\Omega_{xy}|$, which occurs in the topologically nontrivial phases in some systems (including our 2D quantum walk in Sec.~\ref{sec:quantum_walks}), the total area of the manifold coincides with the topological invariant $A={\cal C}/2$ and remains a quantized constant.

We observe that for Dirac models relevant to 2D time-reversal breaking topological insulators\cite{Schnyder08,Ryu10,Chiu16,Chen20191}
\begin{eqnarray}
d_{1}=k_{x},\;\;\;d_{2}=k_{y},\;\;\;d_{3}=M,
\end{eqnarray}
the quantum metric tensor has a meaningful interpretation if we treat the 2D Brillouin zone in momentum space ${\bf k}=(k_{x},k_{y})$ as a manifold. Using Eq.~(\ref{2D_Regab_Imgab}), the quantum metric tensor and the Berry curvature in this Dirac model are given by
\begin{eqnarray}
&&g_{ab}=\left(\begin{array}{cc}
g_{xx} & g_{xy} \\
g_{yx} & g_{yy}
\end{array}
\right)
\nonumber \\
&&=\frac{1}{4\left(M^{2}+k^{2}\right)^{2}}\left(\begin{array}{cc}
k_{y}^{2}+M^{2} & -k_{x}k_{y} \\
-k_{x}k_{y} & k_{x}^{2}+M^{2}
\end{array}
\right),
\nonumber \\
&&\Omega_{xy}=-\Omega_{yx}=\frac{M}{2\left(M^{2}+k^{2}\right)^{3/2}}.
\end{eqnarray}
Moreover, the determinant of the quantum metric tensor coincides with the square of the Berry curvature 
\begin{eqnarray}
\det g_{ab}=\frac{1}{4}\Omega_{xy}^{2}=\frac{M^{2}}{16(M^{2}+k^{2})^{3}}\equiv \chi_{F},
\label{chiF_div_2D2}
\end{eqnarray}
a result very similar to Eq.~(\ref{1D_gkk_Berry2}). To further draw relevance to topological phase transitions driven by the mass term $M$, we see that at the HSP ${\bf k}=(0,0)$, the critical exponents of these quantities are 
\begin{eqnarray}
&&\Omega_{xy}|_{k=0}\propto|M|^{-2}=|M|^{-\gamma},
\nonumber \\
&&\det g_{ab}|_{k=0}\propto|M|^{-4}=|M|^{-2\gamma}.
\end{eqnarray}
Thus the critical exponent of the Berry curvature is basically the same as that of the determinant of the quantum metric tensor. This suggests the determinant $\det g_{ab}\equiv \chi_{F}$ as the representative fidelity susceptibility, and justifies the usage of exponent $\gamma$ for the divergence of the Berry curvature.

In the language of the curvature function in Sec.~\ref{sec:generic_critical_behavior}, our analysis implies that the fidelity is equal to the square of the curvature function (up to a prefactor) for the 1D Dirac model in Sec.~\ref{sec:1D_Dirac_model} and the 2D Dirac model in this section
\begin{eqnarray}
\chi_{F}\propto F({\boldsymbol k},{\boldsymbol M})^{2}.
\end{eqnarray} 
Thus $\chi_{F}$ also takes the Lorentzian shape in Eqs.~(\ref{curv-11D}) and (\ref{curv-22D}), as confirmed by expanding Eqs.~(\ref{chiF_div_1D}) and (\ref{chiF_div_2D2}). As a result, the inverse of the correlation length $\xi^{-1}$ is a momentum scale over which the fidelity susceptibility decays from the HSP.

\section{Quantum walks \label{sec:quantum_walks}}

We now demonstrate that quantum walks serve as practical simulators for the critical exponents, scaling laws, Wannier state correlation functions, CRG, and fidelity ssuceptibility discussed in Sec.~\ref{sec:quantum_criticality_TPT}. The quantum walk is the result of a protocol that has been successively applied on an initial state of a walker. The protocol of the quantum walk consists of two types of operators; coin operators that manipulate the internal states of the walker and shift operators that change the external degree freedom of the walker based on its internal state. Due to successive nature of the quantum walk, the protocol of the quantum walk can be understood as a stroboscopic periodic Floquet evolution. This means that we can map the the protocol of the quantum walk to a (dimensionless) effective Floquet Hamiltonian \cite{Kitagawa}
\begin{eqnarray}
\widehat{H}& = &i \ln\widehat{U}= E \boldsymbol n \cdot \boldsymbol \sigma,  \label{Hamiltonian}
\end{eqnarray}
in which $E$ is the (quasi)energy dispersion, $\boldsymbol \sigma$ are Pauli matrices and $\boldsymbol n$ defines the quantization axis
for the spinor eigenstates at each momentum. It is straightforward to obtain energy through eigenvalues of $\widehat{U}$ by $E= i \ln \eta$, in which $\eta$ is the eigenvalue of $\widehat{U}$. In this paper, we focus on two types of the quantum walk: a) 1D quantum walk with particle-hole (PHS), time reversal (TRS) and chiral (CHS) symmetries in which the symmetries square to $+1$. This type of quantum walk simulates BDI family of the topological phases in one dimension. b) 2D quantum walk with only PHS which simulates D family of topological phases. For both of the protocols, the topological invariants are integer-valued ($\mathbb{Z}$). 
 
The quantum walks have one or two external degrees of freedom (position space) with two internal degrees of freedom. Therefore, the coin Hilbert space ($\mathcal{H}_{C}$) is spanned by $\{ \ketm{0},\: \ketm{1} \}$, respectively. For 1D and 2D quantum walks, the position Hilbert spaces are spanned by $\{ \ketm{x}_{P}: x\in \mathbb{Z}\}$ and $\{ \ketm{x,y}_{P}: x,y \in \mathbb{Z}\}$, respectively. The total Hilbert space of the walk is given by tensor product of subspaces of coin and position spaces. In addition, there are generally three ways that energy bands can close their gap \cite{Panahiyan2020-1}. If the energy bands close their gap linearly, we have a Dirac cone type of boundary state. For the nonlinear case, we have Fermi arc type boundary states. Finally, if the energy bands close their gap for arbitrary momentum, the boundary states are known as flat bands. To summarize, the Dirac cones show linear dispersion while the Fermi arcs have nonlinear dispersive behaviors and the flat bands are dispersionless.

\begin{figure*}[htb]
\begin{center}
\includegraphics[clip=true,width=0.5\columnwidth]{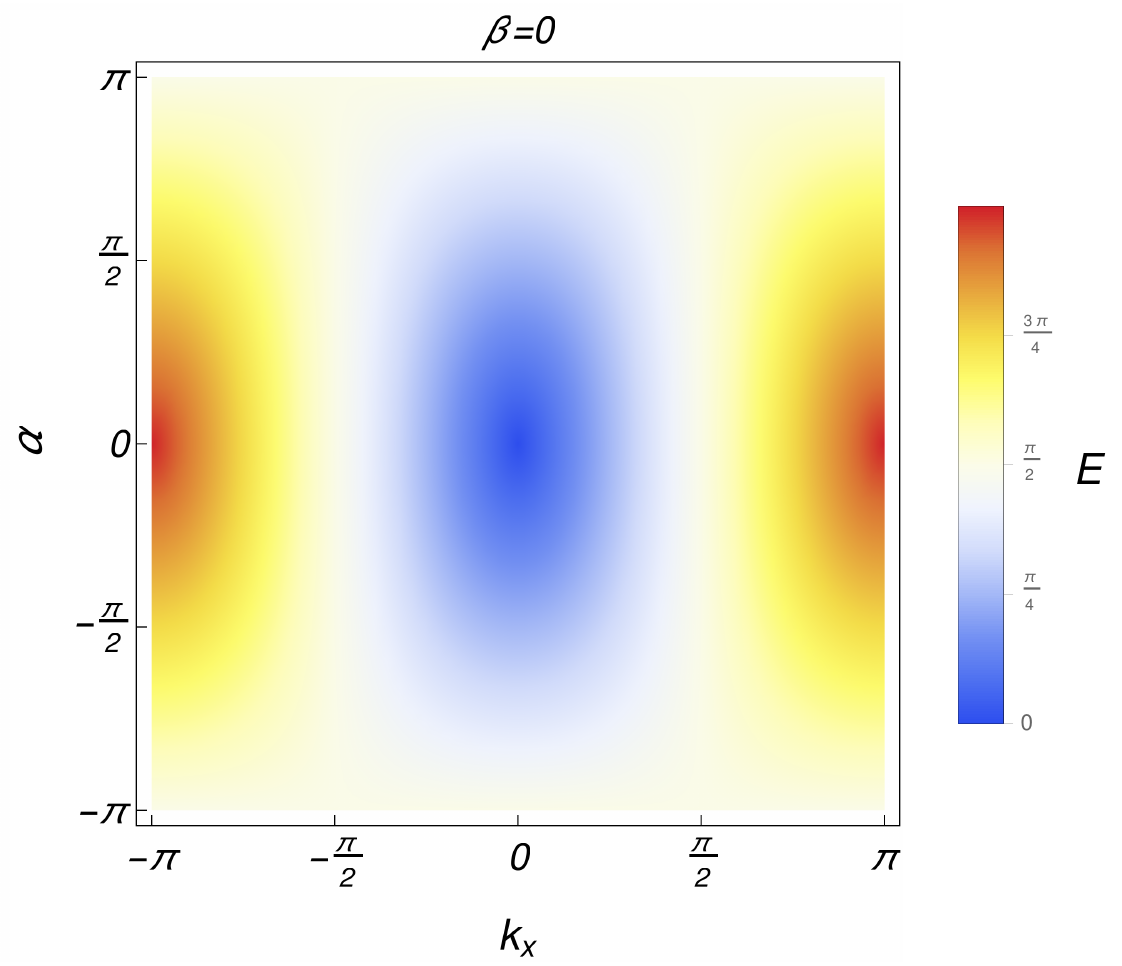}
\includegraphics[clip=true,width=1.2\columnwidth]{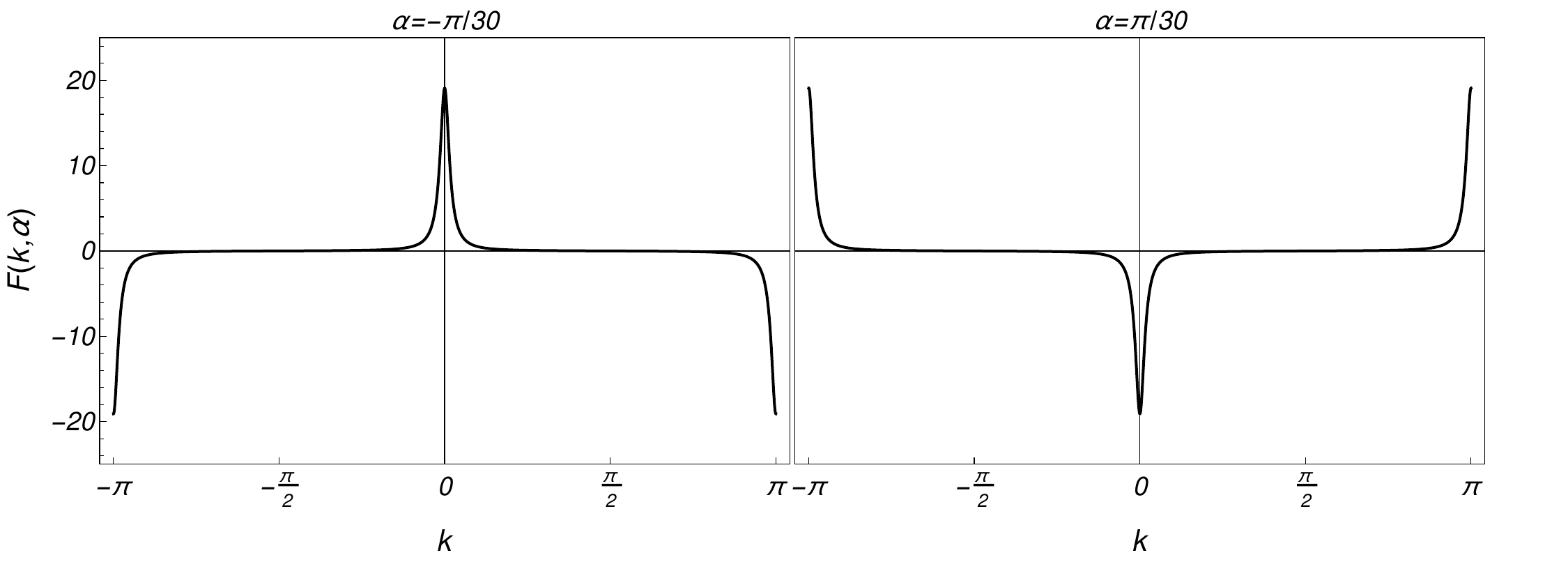}\\
\includegraphics[clip=true,width=0.5\columnwidth]{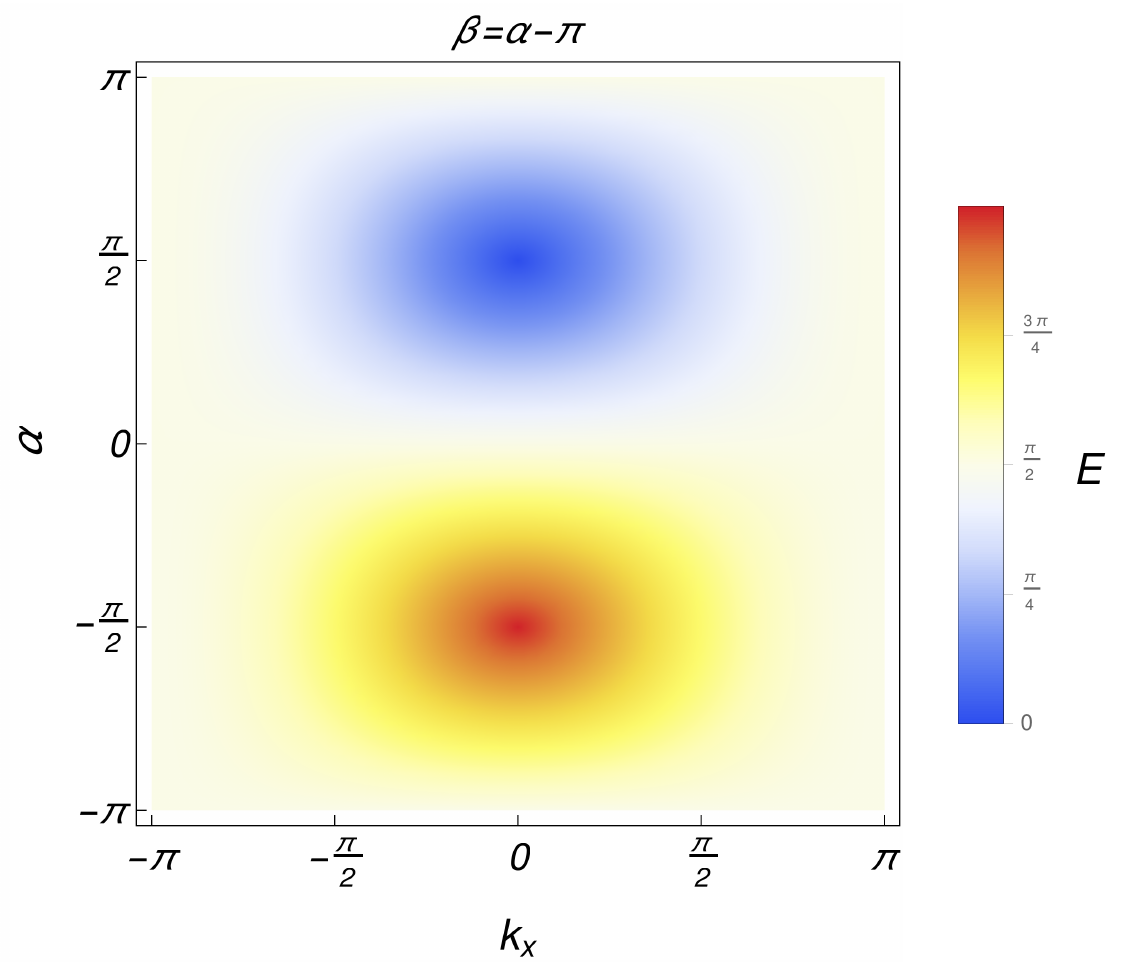}
\includegraphics[clip=true,width=1.2\columnwidth]{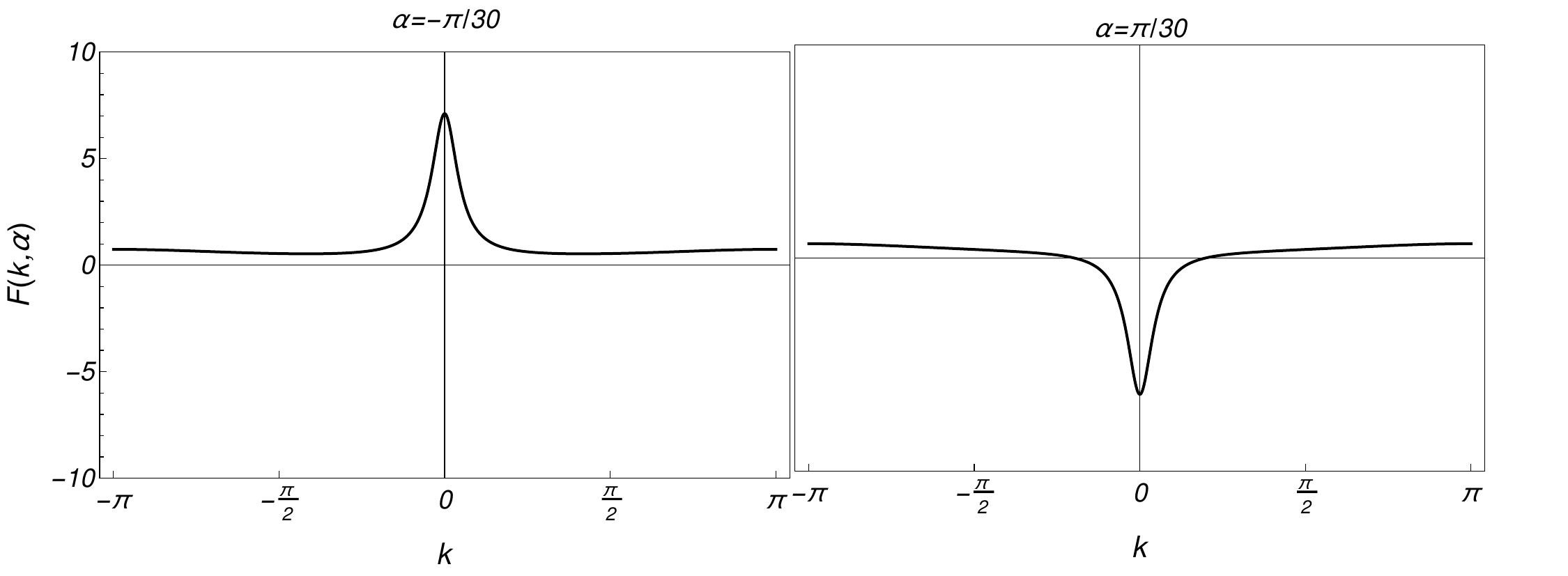}	
   \caption{Left panel, energy as a function of rotation angle and momentum $k$ for two cases of $\beta=0$ and $\beta=\alpha-\pi$ are plotted, respectively. In upper panel, the energy bands close their gap linearly while in lower panel, it is closed nonlinearly. In middle and right panels, the corresponding curvature function as $\alpha \rightarrow \alpha_{c}$ are plotted. Evidently, the curvature function peaks and then flips as it passes the critical point. Since only one peak is present in the diagrams of the curvature function, the band crossing is one.} \label{Fig1}
\end{center}
\end{figure*}

\subsection{One-dimensional class BDI quantum walks}

For the 1D class BDI quantum walks, we consider the following protocol \cite{Kitagawa,Panahiyan2020-1}
\begin{eqnarray}
\widehat{U} & = & \widehat{S}_{\uparrow}(x) \widehat{C}_{y}(\alpha) \widehat{S}_{\downarrow}(x) \widehat{C}_{y}(\beta)  \label{protocol1D},
\end{eqnarray}
in which, one step of quantum walk comprises rotation of internal states with $\widehat{C}_{\beta}$, displacement of its position with $\widehat{S}_{\downarrow}(x)$, a second rotation of internal states with $\widehat{C}_{y}(\alpha)$ and final displacement with $\widehat{S}_{\uparrow}(x)$. The coin operators are rotation matrices around $y$ axis, $\widehat{C}_{y}(\beta)= e^{-\frac{i \beta}{2}\sigma_{y}}$ and $\widehat{C}_{\alpha}= e^{-\frac{i \alpha}{2}\sigma_{y}}$ with $\alpha$ and $\beta$ being rotation angles. The shift operators are in diagonalized forms of $\widehat{S}_{\uparrow}(x)=e^{\frac{i k}{2}(\sigma_{z}-1)}$ and $\widehat{S}_{\downarrow}(x)=e^{\frac{i k}{2}(\sigma_{z}+1)}$ in which we have used Discrete Fourier Transformation ($\ketm{k}=\sum_{x}e^{-\frac{i k x}{2}}\ketm{x}$). It is a matter of calculation to find the energy bands and $\boldsymbol n$ as
\begin{eqnarray}
E & = & \pm\cos^{-1}(\kappa_{\alpha}\kappa_{\beta}\cos(k)-\lambda_{\alpha}\lambda_{\beta}), \label{energy1D}
\end{eqnarray}
\begin{equation}
\boldsymbol n = {\boldsymbol \zeta}/|\zeta|, \label{n1D}
\end{equation} 
in which $\cos(\frac{j}{2})= \kappa_{j}$ and $\sin(\frac{j}{2})= \lambda_{j}$ where $j$ could be $\alpha$ and $\beta$, and $\zeta=(\kappa_{\alpha} \lambda_{\beta} \sin (k),\lambda_{\alpha} \kappa_{\beta} + \kappa_{\alpha} \lambda_{\beta} \sin (k),-\kappa_{\alpha} \kappa_{\beta} \sin (k))$.

The curvature function for the 1D quantum walk can be obtained by \cite{Cardano2017}
\begin{eqnarray}
&& F(k,\alpha,\beta)  =  \bigg (\boldsymbol n \times \partial_{k} \boldsymbol n \bigg) \cdot \boldsymbol A =                                \notag
\\ 
&& \frac{-\cos (k) \lambda_{2\alpha} \kappa_{\beta}-2 \kappa_{\alpha}^2 \lambda_{\beta}}{ 2 \sin ^2(k) \kappa_{\alpha}^2+2\left(\cos (k) \kappa_{\alpha}\lambda_{\beta}+\lambda_{\alpha} \kappa_{\beta}\right)^2}. \label{curv1D}
\end{eqnarray} 
in which $\boldsymbol A=(\kappa_{\beta},0, \lambda_{\beta})$ and perpendicular to $\boldsymbol n$. We consider $\alpha$ as the tuning parameter that closes the band gap, and denote its critical point by $\alpha_{c}$. Using the curvature function in Eq.~\eqref{curv1D}, one can show the following relations
\begin{equation}
\lim\limits_{\alpha \rightarrow \alpha_{c}^{-}} F(k=0,\alpha) = \infty = -\lim\limits_{\alpha \rightarrow \alpha_{c}^{+}} F(k=0,\alpha) ,
\end{equation}
\begin{equation}
\lim\limits_{\alpha \rightarrow \alpha_{c}^{-}} F(k=\pi,\alpha) = - \infty = -\lim\limits_{\alpha \rightarrow \alpha_{c}^{+}} F(k=\pi,\alpha),
\end{equation}
which confirms the divergence and sign change of the curvature function at the critical point described by Eq.~(\ref{FkM_xi_critical_behavior}). To extract the critical exponents and formulate the curvature function, we first gauge away the $z$ component of the ${\boldsymbol\zeta}$ by rotating it around the $y$ axis
\begin{eqnarray}
&&R\,{\boldsymbol\zeta}=\left(\begin{array}{c}
\kappa_{\alpha}\sin (k) \\
\zeta_{y} \\
0
\end{array}\right)=\left(\begin{array}{c}
\zeta_{ x}^\prime \\
\zeta_{ y}^\prime \\
0
\end{array}\right)\equiv{\boldsymbol\zeta}^{\prime},
\nonumber \\
&&R\,{\bf A}=\left(\begin{array}{c}
0 \\
0 \\
1
\end{array}\right)\equiv{\bf A}^{\prime}.
\end{eqnarray}
The new ${\boldsymbol\zeta}^{\prime}$ leads to a rotated Hamiltonian and the corresponding eigenstates
\begin{eqnarray}
|\psi_{k\pm}'\rangle=\frac{1}{\sqrt{2}|{\boldsymbol\zeta}^{\prime}|}\left(\begin{array}{c}
\pm|{\boldsymbol\zeta}^{\prime}| \\
\zeta_{ x}^\prime\pm i\zeta_{ y}^\prime
\end{array}\right),
\end{eqnarray}
in terms of which the curvature function coincides with the stroboscopic Berry connection
\begin{eqnarray}
 && F'(k,\alpha,\beta)= \frac{\zeta_{x}^{\prime}\partial_{k}\zeta_{y}^{\prime}-\zeta_{y}^{\prime}\partial_{k}\zeta_{x}^{\prime}}{(\zeta_{x}^{\prime 2}+\zeta_{y}^{\prime 2})}
=2\langle\psi_{k-}^{\prime}|i\partial_{k}|\psi_{k-}^{\prime}\rangle
\nonumber \\ &&=
\frac{-\kappa_{\alpha}^{2}\lambda_{\beta}-\lambda_{\alpha}\kappa_{\alpha}\kappa_{\beta}\cos (k)}{\kappa_{\alpha}^{2}\sin^{2}(k)+\lambda_{\alpha}^{2}\kappa_{\beta}^{2}
	+2\kappa_{\alpha}\kappa_{\beta}\lambda_{\alpha}\lambda_{\beta}\cos (k)+\kappa_{\alpha}^{2}\lambda_{\beta}^{2}\cos^{2}(k)}.   
\label{strob_Berry_connection}
\nonumber \\
\end{eqnarray}
Notice that because of the controlability of momentum $k$ in the protocol of Eq.~(\ref{protocol1D}), the quantum walk can obtain the entire momentum profile of the Berry connection, and hence the quantum metric in Eq.~(\ref{1D_gkk_Berry2}) for the entire momentum space manifold.

\begin{figure*}[htb]
\begin{center}
\includegraphics[clip=true,width=1\columnwidth]{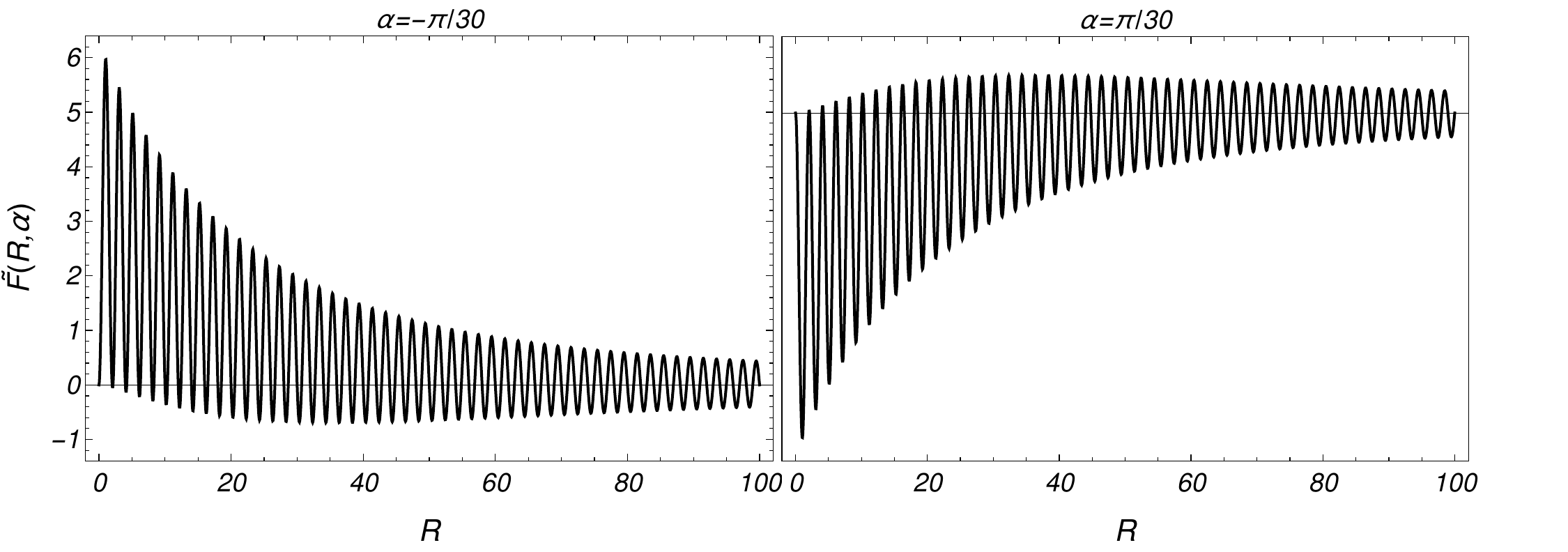}
\includegraphics[clip=true,width=1\columnwidth]{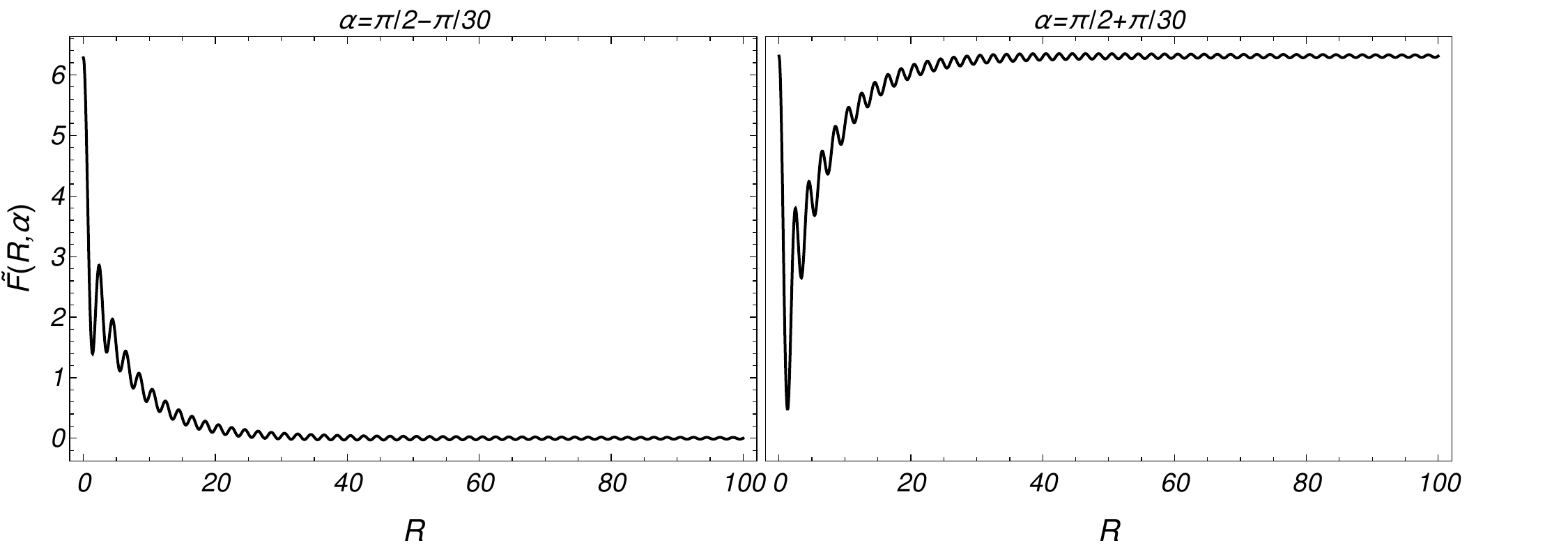}
	\caption{Correlation function, $\tilde{F}_{1D}(R,\alpha)$, as a function of $R$ for two cases of $\beta=0$ (left two panels) and $\beta=\alpha-\pi$ (right two panels). In left two panels, we observe that correlation function decays via a damped oscillation. This oscillation rooted in the fact that curvature function has three peaks at $k=0$ and $\pi$. In contrast, right two panels, the correlation function decays without any oscillation and monotonically. The correlation function, similar to curvature function, also has characteristic behaviors of sign flip that was observed for the curvature function around the critical point.} \label{Fig11}
\end{center}
\end{figure*}

The gap-closing points $k_{c}$ are located at $0$ and $\pi$. Upon a series expansion around these points, we find that the Lorentzian shape in Eq.~(\ref{curv-11D}) satisfy
\begin{eqnarray}
&&F'(k=\left\{0,\pi\right\},\alpha) =- \frac{\kappa_{\alpha}}{\lambda_{\alpha\pm\beta}}\propto\frac{1}{\lambda_{\alpha\pm\beta}}, 
\nonumber \\
&&\xi^{2}(k=\left\{0,\pi\right\},\alpha) =\frac{1}{2}\frac{\kappa_{\beta}^{2}+\kappa_{\alpha}^{2}\kappa_{\beta}^{2}
-\kappa_{\alpha}\kappa_{\beta}\lambda_{\alpha}\lambda_{\beta}}{\lambda_{\alpha\pm\beta}^{2}}
\propto\frac{1}{\lambda_{\alpha\pm\beta}^{2}}. 
\nonumber \\
\label{xiDkc1}
\end{eqnarray} 
In terms of the critical rotation angle $\alpha_{c}$, we find
\begin{eqnarray}
F(k=k_{c},\alpha) \propto 
\xi(k=k_{c},\alpha) \propto  |\alpha-\alpha_{c}|^{-1}.
\end{eqnarray}
which indicate that critical exponents are $\gamma=\nu=1$. This satisfies\cite{Chen2017,Chen19_book_chapter,Chen20191} the 1D scaling law $\gamma=\nu$ and the prediction for 1D class BDI that $\nu\in 2{\mathbb Z}+1$. Finally, using the obtained curvature function, we find that its Fourier transform takes the form 
\begin{eqnarray}
\tilde{F}_{1D}(R,\alpha)\approx\frac{1}{2}\int_{0}^{2\pi}\frac{dk}{2\pi}\frac{F'(k_{c},\alpha,\beta)}{1+\xi^{2}k^{2}} e^{i k R} \propto e^{-R/\xi},\;\;\;\;\;
\label{Fourier_curvature_1D}
\end{eqnarray} 
which decays as a function of $R$ with the length scale $\xi$. Combining with the fact that the curvature function is the stroboscopic Berry connection of the rotated eigenstates, as proved in Eq.~(\ref{strob_Berry_connection}), the Fourier transform in Eq.~(\ref{Fourier_curvature_1D}) then represents the correlation function between the rotated stroboscopic Wannier states as stated in Eq.~(\ref{1D_Wannier_correlation}), with $\xi$ playing the role of the correlation length. Interestingly, if the gap simultaneously closes at $k=0$ and $\pi$, then the correlation function decays through a damped oscillation (see Fig. \ref{Fig11}). On the other hand, if the gap only closes at one momentum, then the correlation function decays monotonically. Finally, the quantum metric associated to the our quantum walk \eqref{1D_gkk_Berry2} and fidelity susceptibility \eqref{1D_fidelity_sus_definition} in one dimension read as 
\begin{eqnarray}
g_{kk}=\chi_{F}= \frac{F^{\prime 2}(k,\alpha,\beta)}{4},
\end{eqnarray} 
which has indication of the divergency provided that $\alpha=\alpha_{c}$ and $k=k_{c}$. 

\begin{figure*}[htb]
\begin{center}
\includegraphics[clip=true,width=0.9\columnwidth]{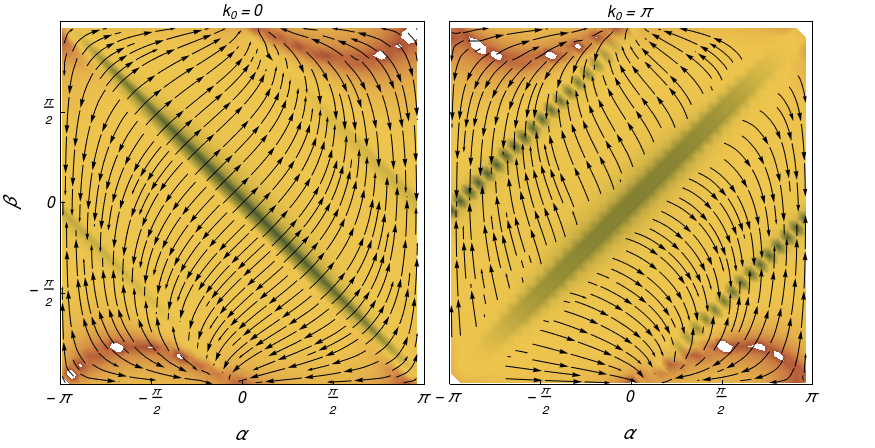}
\includegraphics[clip=true,width=0.51\columnwidth]{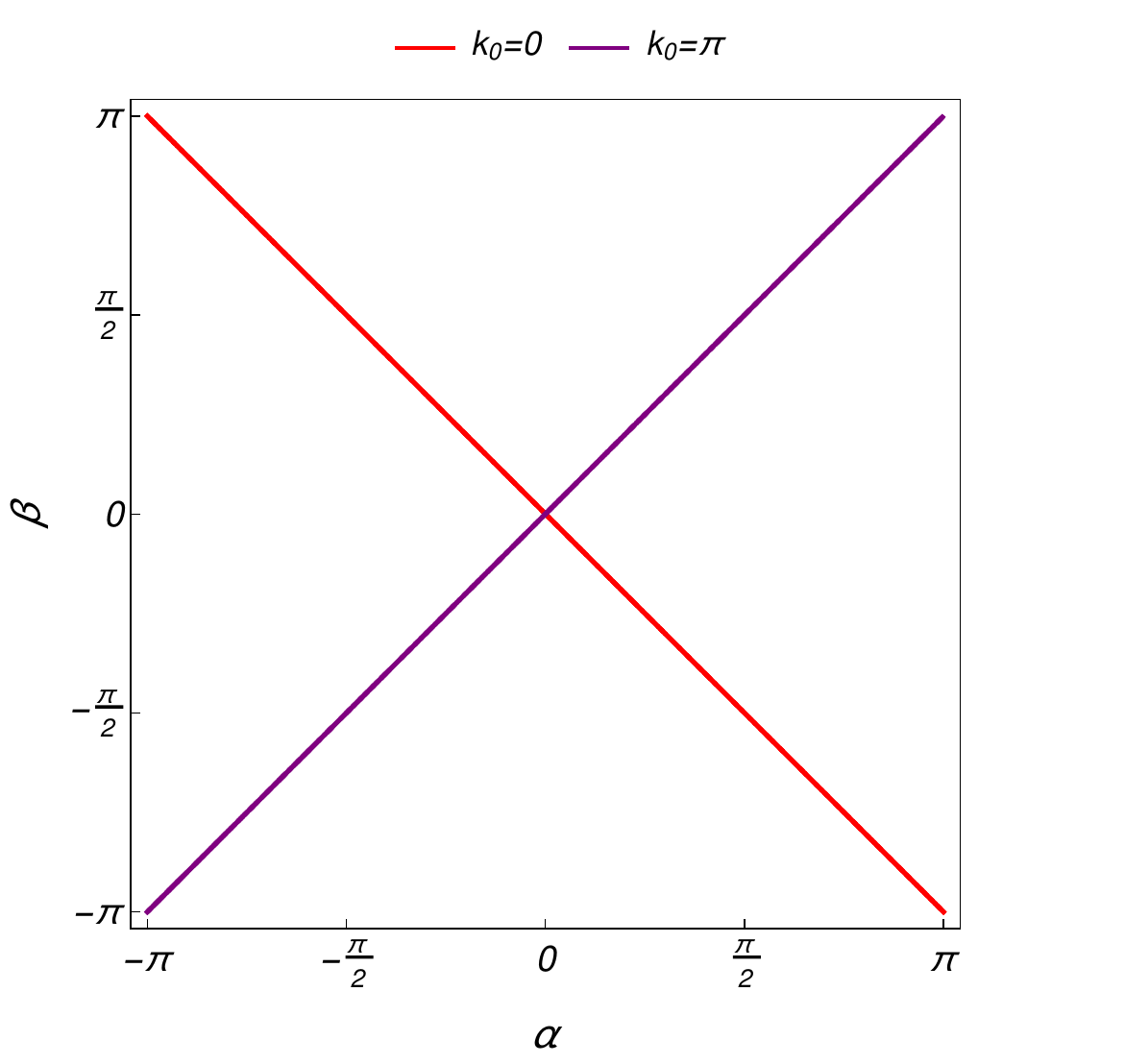}
	\caption{Left two panels, CRG approach applied to the 1D quantum walk shows that the RG flow in the ${\boldsymbol M}=(\alpha,\beta)$ parameter space when the scaling procedure is applied to each of the four HSPs $k_{c}=0$ and $\pi$ with the scaling direction fixed at ${ {\hat k}}_{s}={ {\hat k}}$. The color code indicates the logarithmic of the flow rate $\log|d{\boldsymbol M}/d\ell|$, and the green lines are where the flow rate diverges, indicating a topological phase transition caused by flipping of the curvature function at the corresponding $k_{c}$. Right panel is phase diagrams and critical points as a function of rotation angles. The existence of the multicriticality in the phase diagram is evident.} \label{Fig111}
\end{center}
\end{figure*}

The result of the CRG approach applied to the 1D quantum walk is shown in Fig.~\ref{Fig111}, where we treat ${\boldsymbol M}=\left\{\alpha,\beta\right\}$ as transition-driving parameters. The RG flow is obtained from Eq.~(\ref{RG_eq_numerical}) by using the two HSPs $k_{0}=0$ and $\pi$. The phase boundaries identified from the lines where the RG flow flows away from and the flow rate $|d{\bf M}/d\ell|$ diverges correctly capture the topological phase transitions in this problem, as also confirmed by observing the gap closing in these lines.

\subsection{Two-dimensional class D quantum walks}

\begin{figure*}[htb]
\begin{center}
\includegraphics[clip=true,width=0.5\columnwidth]{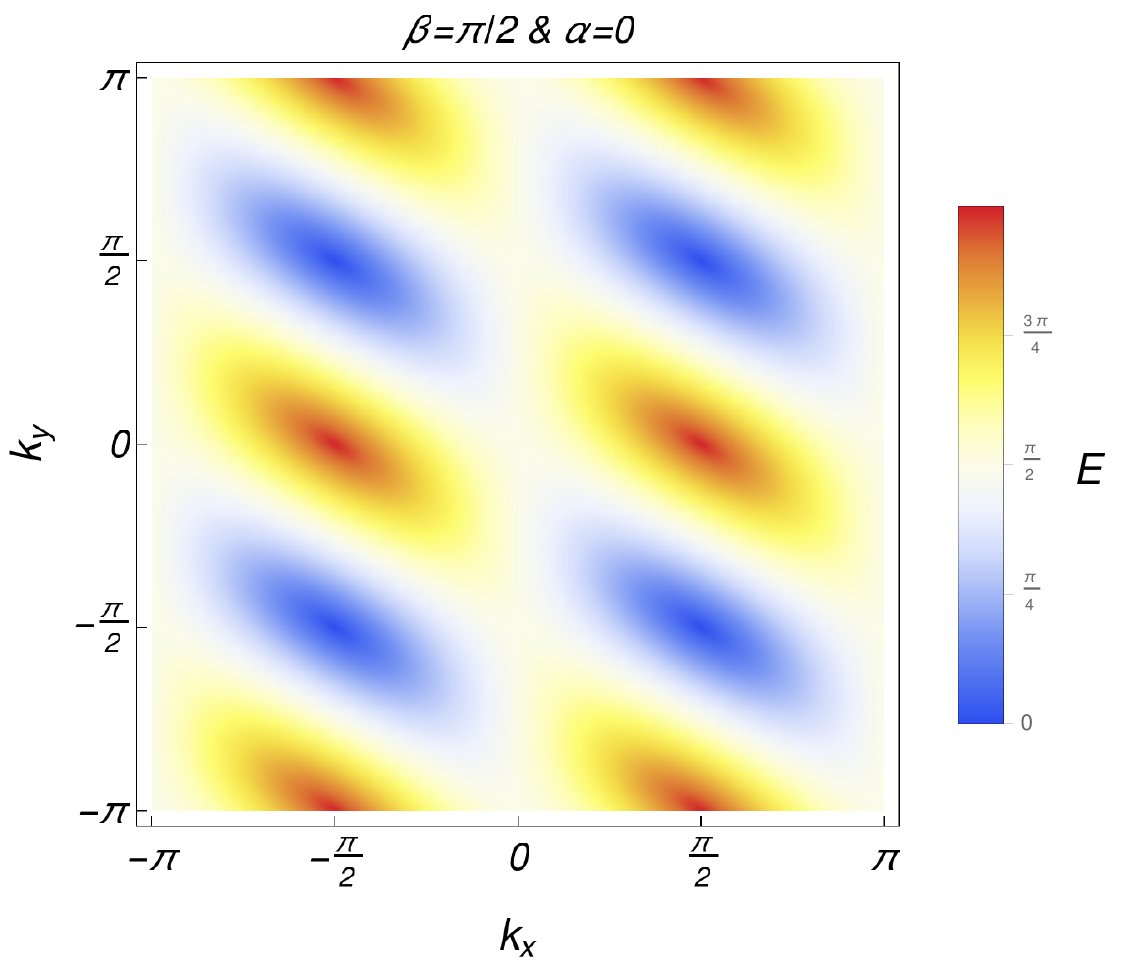}
\includegraphics[clip=true,width=1.2\columnwidth]{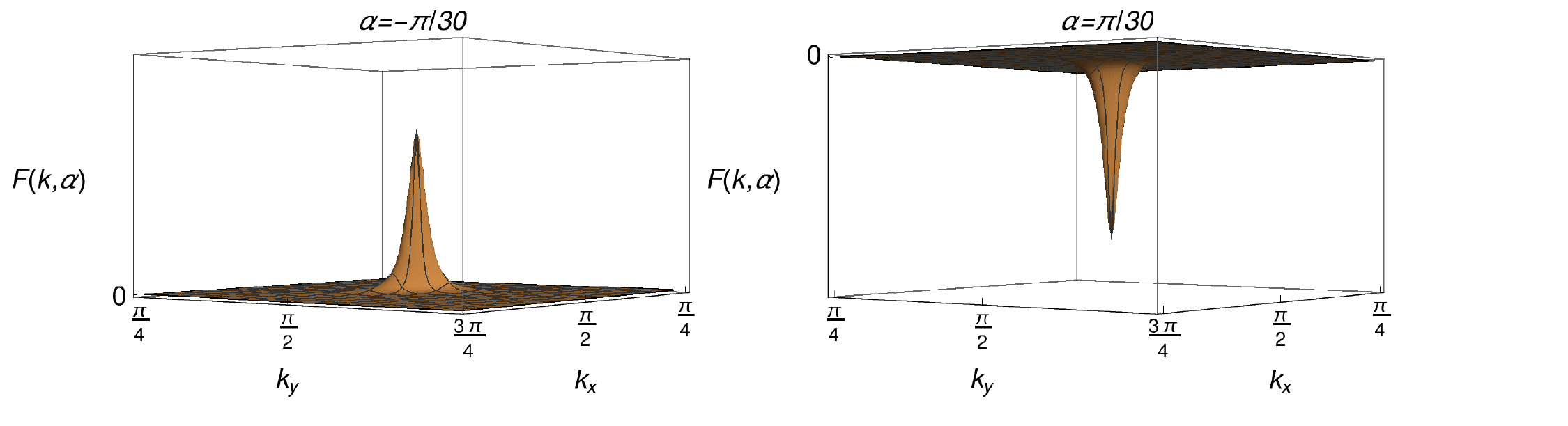}\\
\includegraphics[clip=true,width=0.5\columnwidth]{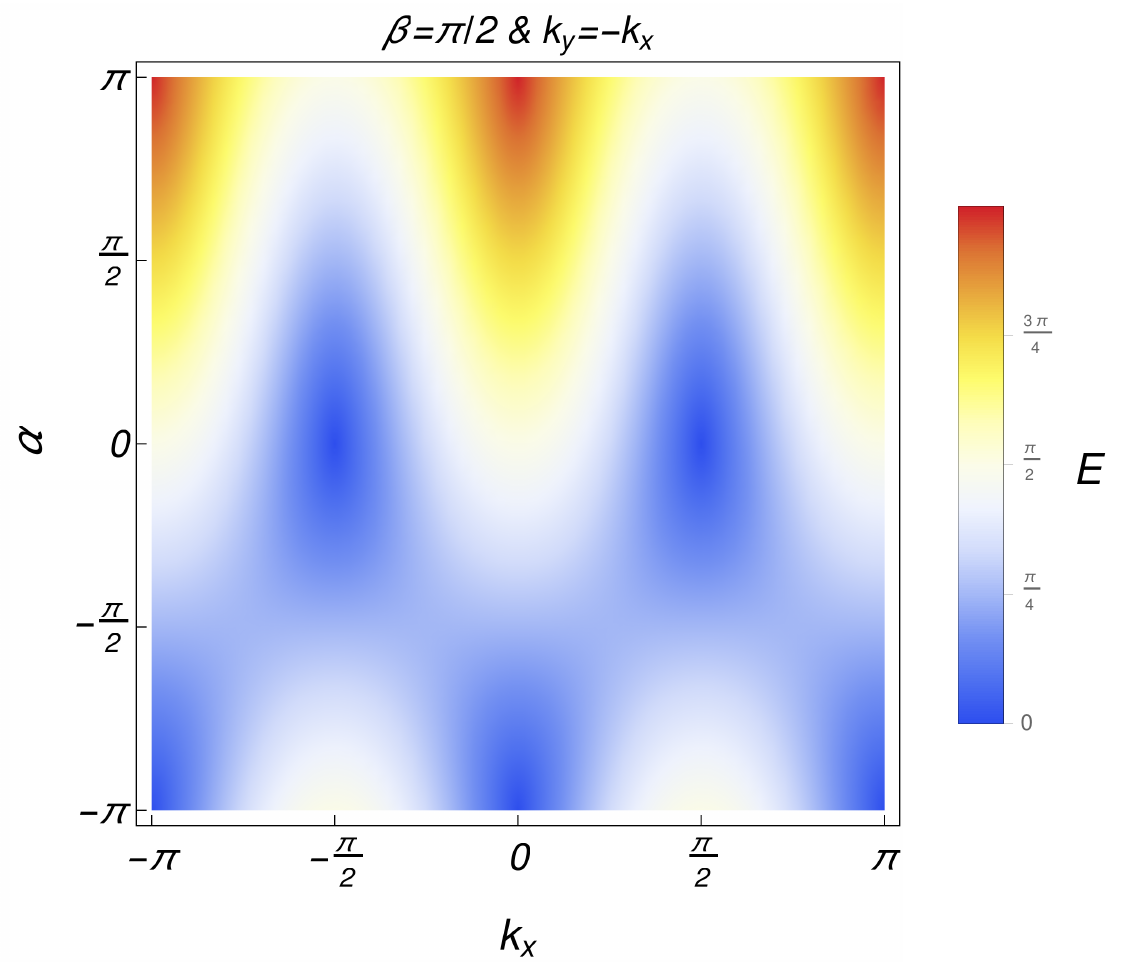}
\includegraphics[clip=true,width=1.2\columnwidth]{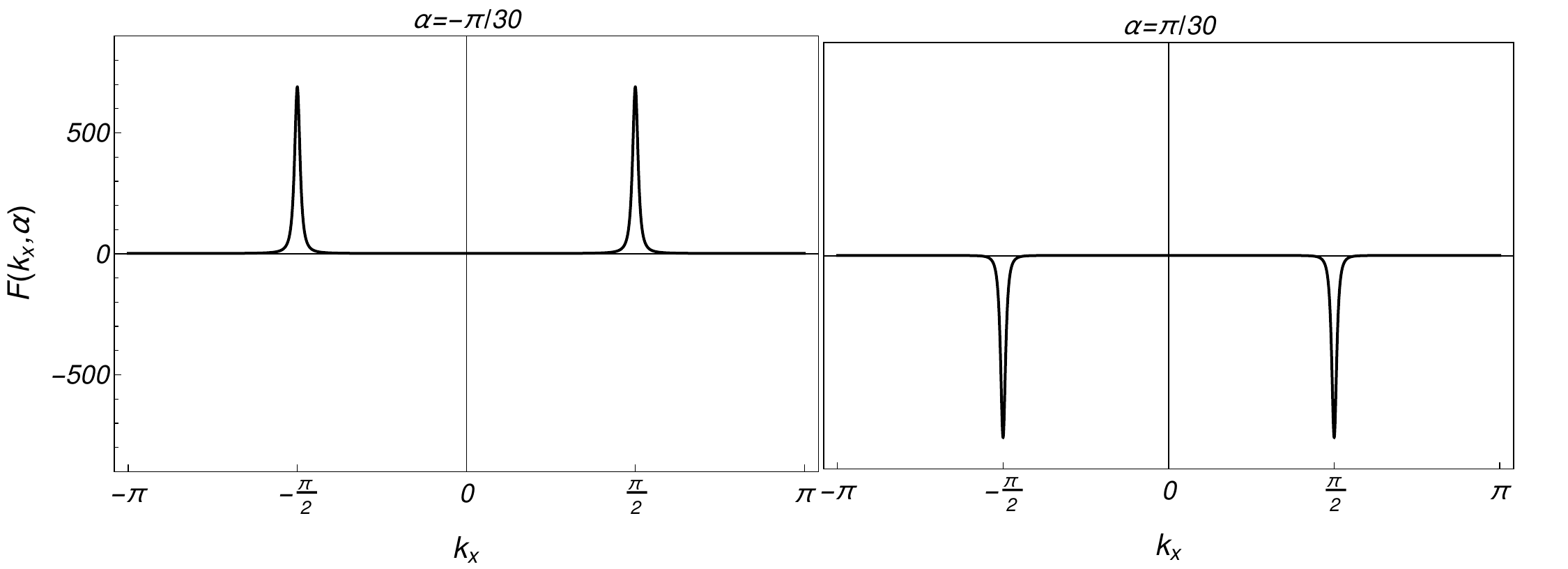}
		\caption{Left panel, energy as a function of rotation angle, momenta $k_{x}$ and $k_{y}$ for two cases of $\beta=\pi/2$ with $\alpha=0$ and $\beta=\pi/2$ with $k_{y}=-k_{x}$ are plotted. Evidently, the curvature function has one peak which indicates that band crossing is one. The peak starts to grow as $\alpha \rightarrow \alpha_{c}$ and it diverges at $\alpha = \alpha_{c}$ with the peak flipping as we pass the critical point.} \label{Fig2}
\end{center}
\end{figure*}

For the 2D class D quantum walk, the protocol of the quantum walk is \cite{Kitagawa,Panahiyan2020-2}
\begin{equation}
\widehat{U} = \widehat{S}_{\uparrow \downarrow}(y) \widehat{C}_{y}(\beta)  \widehat{S}_{\uparrow \downarrow}(x) \widehat{C}_{y}(\alpha) \widehat{S}_{\uparrow \downarrow}(x,y) \widehat{C}_{y} (\beta)  \label{protocol2D},
\end{equation}
Using the Discrete Fourier Transformation, we find the shift operators are $\widehat{S}_{\uparrow \downarrow} (x,y)= e^{i(k_{x}+k_{y}) \sigma_{z}}$, $\widehat{S}_{\uparrow \downarrow} (x)= e^{ik_{x} \sigma_{z}}$ and $\widehat{S}_{\uparrow \downarrow} (y)= e^{ik_{y} \sigma_{z}}$. We obtain the energy bands
\begin{eqnarray}
E & = & \pm\cos^{-1}(\rho), \label{energy2D}
\end{eqnarray}
in which 
\begin{eqnarray}
\rho = &&\kappa_{\alpha}\kappa_{2\beta}\cos(k_x) \cos (k_x+ 2k_y)
\nonumber \\
&&-\kappa_{\alpha}\sin(k_x) \sin (k_x+ 2k_y)
-\lambda_{\alpha} \lambda_{2\beta} \cos ^2(k_x).
\end{eqnarray}
and $\boldsymbol \zeta$ is obtained as
\begin{eqnarray}
\zeta_{x} &=& - 2 \lambda _{\beta } \sin \left(k_x\right) \left(\lambda _{\alpha } \lambda _{\beta } \cos \left(k_x\right)-\kappa _{\alpha } \kappa _{\beta } \cos \left(k_x+2 k_y\right)\right),                   \notag 
\\
\zeta_{y} &=& \lambda _{\alpha } \kappa _{\beta }^2-\lambda _{\alpha } \lambda _{\beta }^2 \cos \left(2 k_x\right)
\nonumber \\
&&+2 \kappa _{\alpha } \kappa _{\beta } \lambda _{\beta } \cos \left(k_x\right) \cos \left(k_x+2 k_y\right),                   \notag     
\\
\zeta_{z} &=& \lambda _{\alpha } \kappa _{\beta } \lambda _{\beta } \sin \left(2 k_x\right)
\nonumber \\
&-&\kappa _{\alpha } \left(\kappa _{\beta }^2 \sin \left(2 \left(k_x+k_y\right)\right)+\lambda _{\beta }^2 \sin \left(2 k_y\right)\right).
\end{eqnarray}
To investigate the critical behavior of the 2D quantum walk, we start from the curvature function
\begin{equation}
F(k_{x},k_{y},\alpha,\beta)= \bigg (\frac{\partial \boldsymbol n}{\partial k_{x}} \times \frac{\partial \boldsymbol n}{\partial k_{y}} \bigg) \cdot \boldsymbol n=\frac{\phi}{(\zeta_{x}^2+\zeta_{y}^2+\zeta_{z}^2)^{\frac{3}{2}}}, \label{Chern}
\end{equation}
whose intergral counts the skyrmion number of the ${\boldsymbol n}$ vector in the BZ, in which
\begin{eqnarray}
\phi=&& 2 \kappa _{\alpha } \lambda _{\beta } \left(\kappa _{\beta }^2+\lambda _{\beta }^2\right) 
\bigg[
4 \kappa _{\alpha }^2 \kappa _{\beta }^2 \lambda _{\beta } \cos \left(k_x\right) \cos \left(k_x+2 k_y\right)+                    \notag
\\ &&
\kappa_{\alpha } \lambda _{\alpha } \kappa _{\beta } \bigg(2 \kappa _{\beta }^2 \cos \left(2 k_y\right) \cos \left(2k_x+2k_y)\right)-                  \notag
\\ &&
\lambda _{\beta }^2 \left(2 \cos \left(2 k_x\right)+\cos \left(4
k_y\right)+3\right)\bigg)                  \notag
\\ &&
+2 \lambda _{\alpha }^2 \lambda _{\beta } \cos \left(2 k_y\right) \left(\lambda _{\beta }^2-\kappa _{\beta }^2 \cos \left(2 k_x\right)\right)
\bigg].                  \notag
\end{eqnarray}
The controlability of momentum ${\boldsymbol k}$ in the protocol of Eq.~(\ref{protocol2D}) enables us to obtain the entire momentum profile of the Berry curvature, and hence the fidelity susceptibility in Eq.~(\ref{chiF_div_2D}) for the entire momentum space manifold. Now, we consider the rotation angle $\alpha$ as the tuning parameter. In contrast to the 1D case, the energy bands can close their gap at different values of $k_{x},k_{y}$ not limited to $0$ and $\pi$. Nevertheless, irrespective of the precise location of ${\boldsymbol k}_{c}$, the divergence and flipping of the curvature function always hold
\begin{eqnarray}
&&\lim\limits_{\alpha \rightarrow \alpha_{c}^{-}} F(k_{x}=k_{c},k_{y}=k_{c},\alpha) 
\nonumber \\
&&= -\lim\limits_{\alpha \rightarrow \alpha_{c}^{+}} F(k_{x}=k_{c},k_{y}=k_{c},\alpha)= \pm \infty,
\end{eqnarray}
indicating that the curvature function can be invoked to study the critical behavior of the system. The plotted diagrams for curvature function show the emergence of a single peak as $\alpha \rightarrow \alpha_{c}$ (see Fig. \ref{Fig2}). This shows that band crossing is one ($n=1$) and peak-divergence scenario is applicable for this protocol as well. In what follows, for the sake of brevity and simplicity, we consider $k_{y}=-k_{x}$. In such a case, one can find the energy gap closes at $k_{x}=k_{c}=\pi/2$. It is straightforward to find the curvature function and the length scale at the critical points as 
\begin{eqnarray}
&&F(k_{x}=-k_{y}=\frac{\pi}{2},\alpha) =\frac{2 {\rm Sig}(\kappa_{\alpha}) \left(\lambda_{2(\alpha-\beta)}-\lambda_{2\alpha}-\lambda_{2\beta} \right) }{1- \kappa_{2\alpha}},
\nonumber \\
&&\xi_{x}^2(k_{x}=-k_{y}=\frac{\pi}{2},\alpha) \approx \frac{\Xi}{(1- \kappa_{2\alpha})^{\frac{1}{2}}}, \label{xiDkc11}
\end{eqnarray} 
in which 
\begin{eqnarray}
\Xi = &&
2\sqrt{2} \kappa_{\alpha} (2\lambda_{\beta}^2(5+2 \kappa_{2\alpha}+\kappa_{2\beta})
\nonumber \\
&&-\lambda_{2\beta}\cot \left(\frac{\alpha
}{2}\right) (3\kappa_{2\beta}+\kappa_{2\alpha}-4)). \notag
\end{eqnarray}
By setting $\beta=\pi/2$, critical point occurs at $\alpha_{c}=0$, in which case we extract the critical exponents $\gamma=2$ and $\nu=1$, and the scaling law is valid through $\gamma= D\nu$ with $D=2$ which is in agreement with the results in Ref. \cite{Chen20191,Molignini19}.

\begin{figure*}[htb]
\begin{center}
\includegraphics[clip=true,width=1.2\columnwidth]{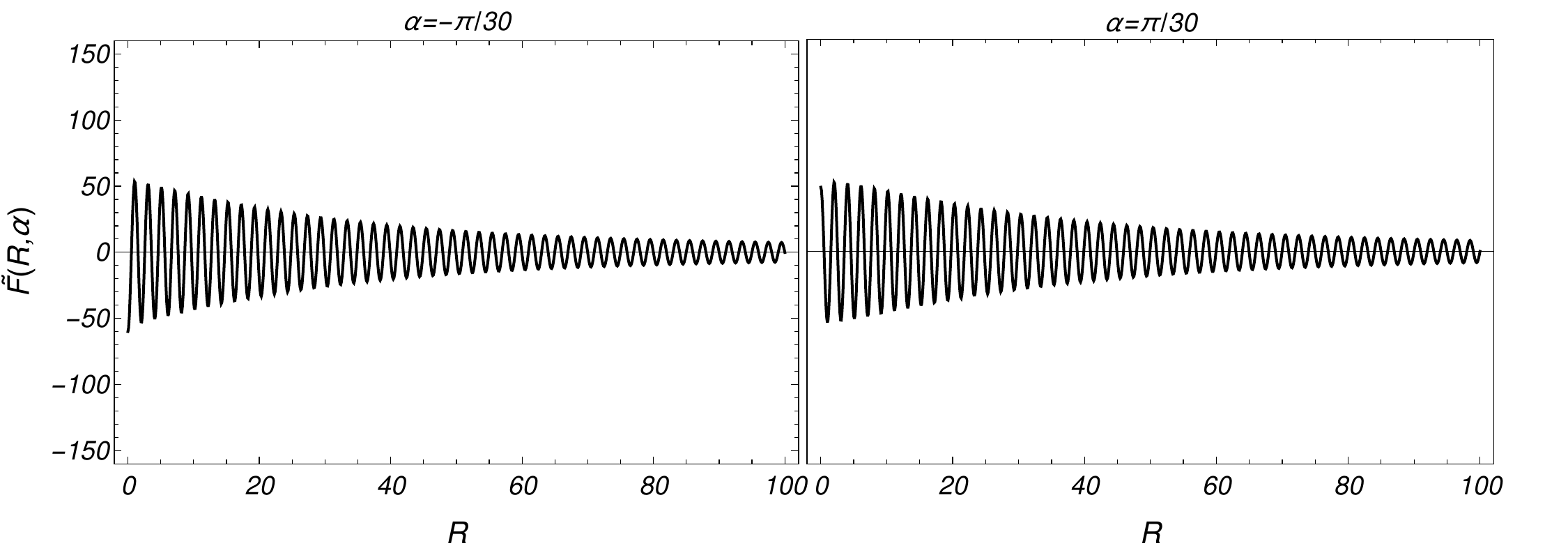}
	\caption{Correlation function, $\tilde{F}_{2D}(\boldsymbol R,\alpha)$, as a function of $\boldsymbol R$ for $\beta=\pi/2$ with $k_{y}=-k_{x}$ ($R_{y}=-R_{x}$). The correlation function decays through a damped oscillation since the curvature function acquires two peaks. Evidently, the correlation function, similar to cuvrature function, flips sign as system passes the critical point. } \label{Fig22}
\end{center}
\end{figure*}

To see the correlation function, we start from the stroboscopic eigenstates of the Hamiltonian
\begin{eqnarray}
|\psi_{\bf k\pm}\rangle=\frac{1}{\sqrt{2n(n\pm n_{z})}}\left(\begin{array}{c}
n_{z}\pm n \\
n_{x}+i n_{y}
\end{array}\right).
\end{eqnarray}
from which we see that the stroboscopic Berry Curvature of the filled band eigenstates coincides with the curvature function in Eq.~(\ref{Chern})
\begin{eqnarray*}
\partial_{k_{x}}\langle\psi_{\boldsymbol k-}|\partial_{k_{y}}|\psi_{\boldsymbol k-}\rangle-
\partial_{k_{y}}\langle\psi_{\boldsymbol k-}|\partial_{k_{x}}|\psi_{\boldsymbol k-}\rangle=\frac{1}{2}F({\boldsymbol k},\alpha,\beta),
\end{eqnarray*} 
Thus the Fourier transform of the curvature function gives a correlation function that measures the overlap of the stroboscopic Wannier states according to Eq.~(\ref{Wannier_correlation_2D}). Moreover, using the Lorentzian shape in Eq.~(\ref{curv-22D}), the Fourier transform
\begin{eqnarray}
\tilde{F}_{2D}(\boldsymbol R,\alpha) &\approx&\frac{1}{2}\int\frac{d^{2}{\boldsymbol k}}{(2\pi)^{2}}\frac{F(\boldsymbol k_{c},\alpha)}{1+\xi^{2}\delta \boldsymbol k^{2}},
\end{eqnarray}
gives a correlation function that decays with ${\boldsymbol R}$ with the correlation length $\xi$. In case of $k_{y}=-k_{x}$ ($R_{y}=-R_{x}$), we observe that decay of the correlation function would be through a damped oscillation which is due to presence of at least two peaks in curvature function (see Fig. \ref{Fig22}). Now, we are in a position to find the fidelity susceptibility which can be done by first building up the quantum metric associated to the our quantum walk \eqref{2D_Regab_Imgab} and then calculating its determinant which leads to 
\begin{eqnarray}
\chi_{F}=\text{det } g_{k_{x}k_{y}}= \frac{F^{2}(k_{x},k_{y},\alpha,\beta)}{4},
\end{eqnarray} 
which has divergency at gapless points of energy bands and can be used to characterize the critical points. 

\begin{figure*}[htb]
\begin{center}
\includegraphics[clip=true,width=1.5\columnwidth]{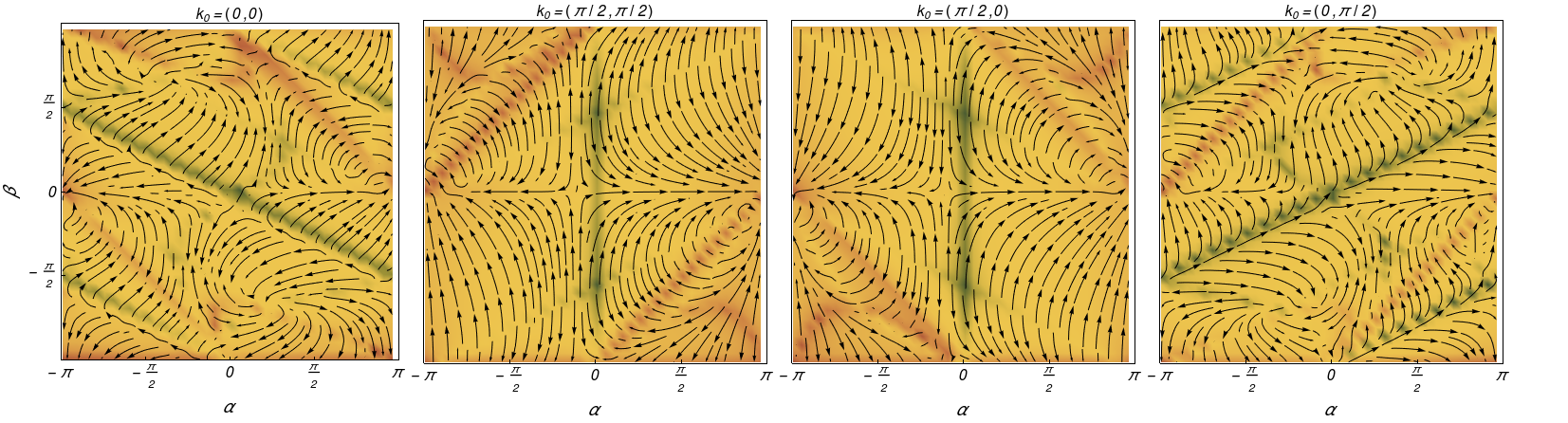}
\includegraphics[clip=true,width=0.34\columnwidth]{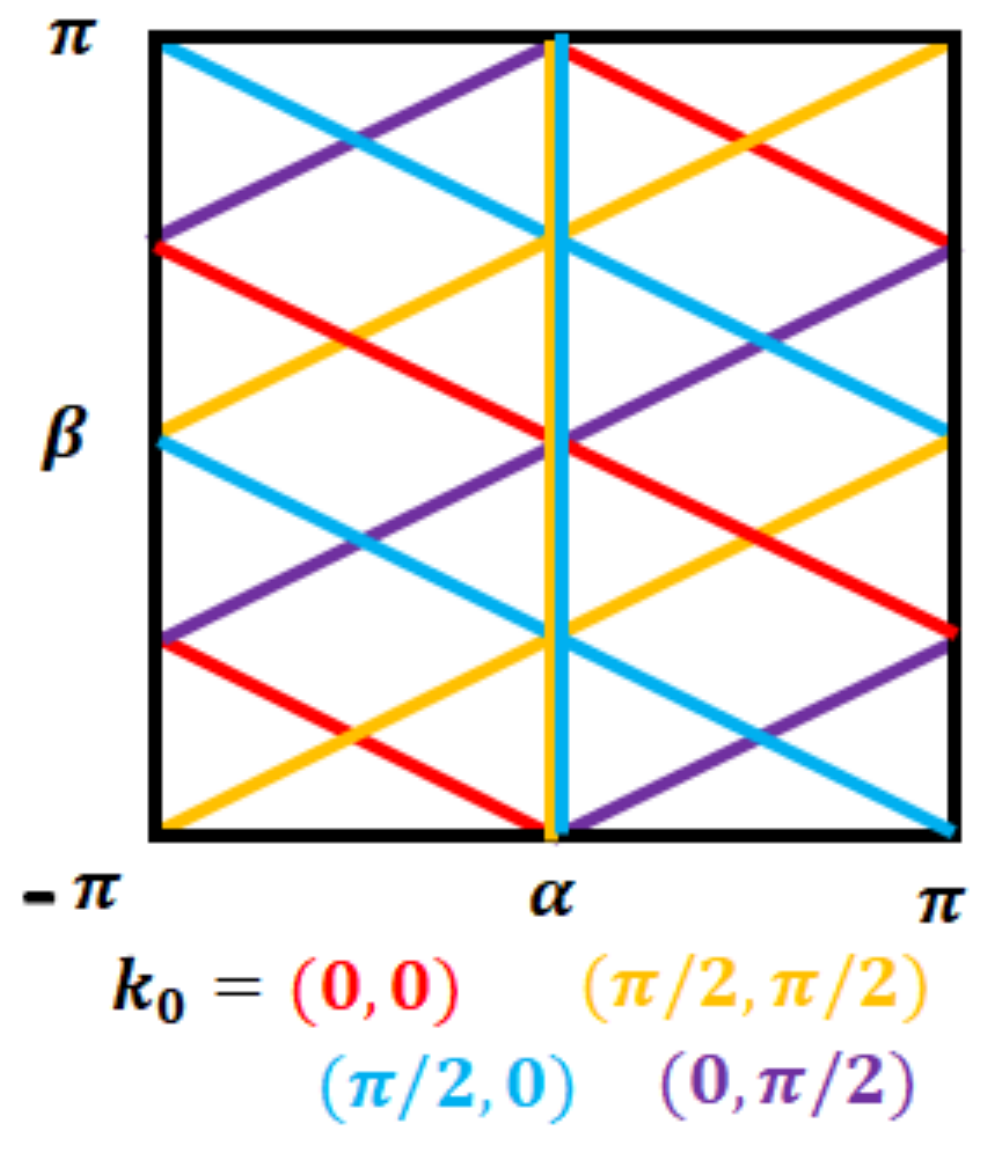}
	\caption{CRG approach applied to the 2D quantum walk, where the left four panels show the RG flow in the ${\boldsymbol M}=(\alpha,\beta)$ parameter space when the scaling procedure is applied to each of the four HSPs ${\boldsymbol k}_{0}=(0,0)$, $(\pi/2,\pi/2)$, $(\pi/2,0)$, and $(0,\pi/2)$ with the scaling direction fixed at ${\boldsymbol {\hat k}}_{s}={\bf {\hat k}}_{x}$. The color code indicates the logarithmic of the flow rate $\log|d{\boldsymbol M}/d\ell|$, and the green lines are where the flow rate diverges, indicating a topological phase transition caused by flipping of the curvature function at the corresponding ${\boldsymbol k}_{0}$. The right panel shows the phase boundaries by combining these results. } \label{Fig222}
\end{center}
\end{figure*}

We now discuss the application of the CRG approach to the 2D quantum walk, treating ${\boldsymbol M}=\left\{\alpha,\beta\right\}$ as a 2D parameter space. The resulting RG flow is shown in Fig. \ref{Fig222}, where we use the four HSPs ${\boldsymbol k}_{0}=(0,0)$, $(\pi/2,\pi/2)$, $(\pi/2,0)$ and $(0,\pi/2)$ and fix the scaling direction to be ${\bf {\hat k}}_{s}={\bf {\hat k}}_{x}$. The lines in the parameter space where the flow rate $|d{\bf M}/d\ell|$ diverges and the RG flow flows away from are the topological phase transitions caused by flipping the curvature function at the corresponding ${\bf k}_{0}$. Collecting the transition lines in all the four ${\bf k}_{0}$ cases correctly captures the phase diagrams for 2D quantum walk.

\section{Conclusion \label{sec:conclusion}}

In summary, we clarified the notion of quantum metric tensor near topological phase transitions within the context of 1D and 2D Dirac models. The quantum metric tensor defined in the manifold of momentum space turns out to represent a kind of geometric texture of the ${\hat{\bf d}}$-vector that parametrizes the Dirac Hamiltonian, in a way similar to the winding texture of Berry curvature and the skyrmion texture of Berry curvature. The determinant of the quantum metric tensor coincides with the square of the Berry connection in 1D and Berry curvature in 2D, from which we define the representative fidelity susceptibility. As a result, the fidelity susceptibility shares the same Lorentzian shape and critical exponent as these curvature functions, and moreover the correlation length yields a momentum scale over which the fidelity susceptibility decays. 

We then turned to the simuation of these quantities by means of quantum walks for 1D class BDI and 2D class D Dirac models. It is shown that not only can the quantum walks map out the entire momentum profile of the curvature function, and hence the quantum metric tensor in the entire manifold, but also capture the critical exponents and scaling laws. Due to geometry of the gap-closing in 1D quantum walk, the stroboscopic Wannier state correlation function either displays a damped oscillation which happens for Dirac cone gap-closings, or monotonically decays for the Fermi arc case. For 2D quantum walk, since the curvature function admitted presence of two peaks corresponding to two critical points, the correlation function decayed through a damped oscillation. These results confirm the quantum walks as universal simulators of topological phase transitions, and introduces the notion of universality class into these simulators that can eventually be compared with real topological materials.

While the present work focuses on only two protocols of quantum walks, the same analysis of criticality can be done for protocols that simulate other symmetry classes and dimensions. In addition, the decay of the correlation function for robust edge states observed in inhomogenous quantum walk is another subject of the interest \cite{Kitagawa}. From the perspective of Foquet engineering, it remains to be explored how to properly design the protocol such that the quantum walk can simulate more exotic topological phases, such as nodal loops\cite{Molignini20}. On the fidelity susceptibility side, it remains to be analized how other kinds of curvature functions, such as that associated with the ${\mathbb Z}_{2}$ invariant and the 3D winding number\cite{Chen2,Chen19_book_chapter,Chen20191}, are related to the quantum metric tensor. We leave these intriguing issues for the future investigations.

\section{Acknowledgement}

The authors acknowledge fruitful discussions with P. Molignini, R. Chitra and S. H. Hendi. W. Chen is financially supported by the productivity in research fellowship from CNPq.

\bibliography{liter}

\end{document}